\documentclass{article}

\usepackage{arxiv}

\usepackage[utf8]{inputenc} % allow utf-8 input
\usepackage[T1]{fontenc}    % use 8-bit T1 fonts
\usepackage{hyperref}       % hyperlinks
\usepackage{url}            % simple URL typesetting
\usepackage{amsfonts}       % blackboard math symbols
\usepackage{nicefrac}       % compact symbols for 1/2, etc.
\usepackage{microtype}      % microtypography
\usepackage{multirow}
\usepackage{tablefootnote}
\usepackage{footnote}
\usepackage{graphicx}
\usepackage{multirow}
\usepackage{afterpage}
\usepackage[numbers,sort&compress]{natbib}
\usepackage{fullpage}
\usepackage{caption}
\usepackage{needspace}
\usepackage{times}
\usepackage{fancyhdr,graphicx,amsmath,amssymb}
\usepackage{multirow}
\usepackage{graphicx}

\title{Machine learning based automated identification of thunderstorms from anemometric records using shapelet transform}

\author{
  Monica Arul\\
  NatHaz Modeling Laboratory\\
  Department of Civil Engineering\\
  University of Notre Dame\\
  Notre Dame, IN 46556 \\
  \texttt{maruljay@nd.edu} \\
   \And
 Ahsan Kareem \\
  NatHaz Modeling Laboratory\\
  Department of Civil Engineering\\
 University of Notre Dame \\
  Notre Dame, IN 46556 \\
  \texttt{kareem@nd.edu} \\
  \AND
  Massimiliano Burlando  \\
  University of Genoa \\
  Via Balbi, 5, 16126, Genoa, Italy \\
  \texttt{massimiliano.burlando@unige.it} \\
  \And
  Giovanni Solari  \\
University of Genoa \\
  Via Balbi, 5, 16126, Genoa, Italy \\
  \texttt{giovanni.solari@unige.it} \\
}

\begin{document}
\maketitle

\begin{abstract}
Detection of thunderstorms is important to the wind hazard community to better understand extreme wind field characteristics and associated wind-induced load effects on structures. This paper contributes to this effort by proposing an innovative course of research that uses machine learning techniques, independent of wind statistics-based parameters, to autonomously identify thunderstorms from large databases containing high-frequency sampled continuous wind speed data. In this context, the use of Shapelet transform is proposed to identify key individual attributes distinctive to extreme wind events based on similarity of the shape of their time series signature. This shape-based representation, when combined with machine learning algorithms, yields a practical event detection procedure with minimal domain expertise. In this paper, the shapelet transform along with Random Forest classifier is employed for the identification of thunderstorms from 1-year of data from 14 ultrasonic anemometers that are a part of an extensive in-situ wind monitoring network in the Northern Mediterranean ports. A collective total of 240 non-stationary records associated with thunderstorms were identified using this method. The results lead to enhancing the pool of thunderstorm data for a more comprehensive understanding of a wide variety of thunderstorms that have not been previously detected using conventional gust factor-based methods.
\end{abstract}

% keywords can be removed

\section{INTRODUCTION}
A modern study of thunderstorms was first accomplished by the Thunderstorm Project \cite{byers1949thunderstorm} that was commissioned by the U.S. Congress in 1945 after a series of severe thunderstorm-related aircraft incidents. The study proved that thunderstorms are mesoscale phenomena that develop over a few kilometers and consist of convective cells that evolve through three stages (cumulus, mature and dissipating stages) in which an updraft of warm air is followed by a downdraft of cold air.The report rendered a preliminary understanding of the life cycle, form, and distribution of thunderstorms and their associated physical processes, describing for the first time the evolution of the so-called single-cell thunderstorm as the basic organizational structure of all convective thunderstorms. These studies facilitated the burgeoning growth of research in the field of atmospheric sciences that used anemometers, radars, barometers, radiosondes, and instrumented aircraft to study more about the causes and attributes of thunderstorm-induced winds. \cite{browning1964airflow} used the single-cell idea to describe a special form of deep convection that was later named supercell \cite{browning1977structure}. Multicell-thunderstorms were defined as systems where new cells are initiated along the outflow boundary spreading out from previous cells \cite{weisman1982dependence,fovell1995temporal}. Based on the multi-cell idea, larger (mesoscale to synoptic scale) patterns of thunderstorms were also discovered that organized along straight lines of convection and were called squall lines \cite{house1959mechanics,weisman1988structure}.

\cite{charba1974application} used the data from anemometers, thermometers, barometers and hygrometers mounted along a 444 m high transmission tower to study an intense event that occurred in Oklahoma in May 1969. \cite{craig1976vertical} analyzed the characteristics of 20 thunderstorm outflows using radar images combined with the data collected from a 461 m tower equipped with meteorological sensors. \cite{wakimoto1982life} carried out a detailed analysis on the life cycle of thunderstorm gust fronts using the data collected from Doppler radars, mesonets stations and rawinsondes. \cite{sherman1987passage} described the passing of a thunderstorm downburst over a suburb of Brisbane, Australia with the help of data collected from an instrumented tower along with radar images. \cite{gast2003comparison} documented the varying thermodynamic and kinematic characteristics of two extreme thunderstorm outflows utilizing high-resolution near-surface observations. \cite{gast2003supercell} analyzed a super-cell thunderstorm that produced a rear-flank downdraft which passed over 7 monitored towers in Lubbock, Texas, in June 2002; they also examined data from doppler radars and meteorological soundings. \cite{gunter2015high} provided a deeper understanding of the characteristics of thunderstorm outflow winds and wind profiles using data collected for Project SCOUT using mobile Doppler radars. 

During the same period when thunderstorms were researched with fervor by the meteorologists, the wind engineers also realized that several wind events in the mid-latitude region that caused catastrophic damages to the built environment were mainly due to thunderstorms \cite{letchford2002thunderstorms}. This led to a surge in research in the field of wind engineering guided by the one that took place in atmospheric science. The wind engineering line of research is mostly based on anemometric recordings, without analyzing the meteorological information related to the wind event or the weather scenarios out of which they developed or, at most, utilizing the correlation between some meteorological quantities or phenomena (e.g., rain, surface temperature, thunder, lightning) to infer the presence of thunderstorm cells and strong convection in the atmosphere. A majority of the initial research focused on identifying thunderstorms from a mixed-wind climate to enable separate analyses \cite{gomes1978extreme}. \cite{gomes1976thunderstorm} studied the separation of thunderstorms and non-thunderstorms and also obtained distributions for gust speed wind speeds in Australia. \cite{riera1989pilot} segregated thunderstorms from other extreme wind events using wind duration, presence of thunder or lightning, rainfall, and sudden fall in the temperature. \cite{twisdale1992research} examined the statistics of thunderstorm and non-thunderstorm winds collected from various cities in the United States. \cite{choi1999extreme} classified wind events as thunderstorms and non-thunderstorms based on whether thunder and rain were recorded. \cite{choi2002gust} investigated the variation of gust factors during thunderstorms and monsoon winds and found that the gust factor of thunderstorms is always higher when compared to the monsoon winds.  \cite{kasperski2002new} claimed that in mid-latitude areas, thunderstorms cannot be clearly separated from synoptic depressions due to the existence of a third set of events, called gust fronts, which present intermediate properties. His criterion of extraction was based on the evaluation of three parameters: mean wind speed, peak wind speed and gust factor. An automated method to extract and classify thunderstorms from non-thunderstorm wind data present in the Automated Surface Observing System (ASOS) was developed by \cite{lombardo2009automated} for extreme value analysis. The identification of thunderstorms was based on weather observations, peak wind data and thunderstorm start and end times reported by manual observers. \cite{de2014separation} established a semi-automated method based on gust factor to separate and classify extra-tropical depressions, thunderstorms, and gust fronts using quantitative controls and qualitative judgments. During these developments, concurrently models for simulation of downburst events \cite{mason2010numerical} and for capturing the load effects of thunderstorm outflows have been advanced, which can significantly benefit from the additional knowledge derived from new data being collected \cite{kwon2009gust,solari2020detection}. Also, a state of the art in non-synoptic wind storm research focusing on downburst monitoring, modeling, detections, effects of anthropogenic climate change can be found in \cite{hangan2021oxford}.

The meteorological and wind engineering fields have witnessed over the years, a tremendous increase in the volume of data related to wind events due to the rapidly growing wind monitoring networks and stations that can continuously record wind field measurements with high sampling rates. This has led to massive amounts of high-dimensional data collected continuously over time and stored as a time series. From a “big data” perspective, many of the meteorological and wind engineering techniques mentioned above, hold little utility for mining massive wind databases that require indexing, predicting, classifying, clustering, segmenting, and identifying patterns from data. Thus, given the wide prevalence of big data, there has been increased attention in using machine learning techniques to automatically detect desired events from large databases \cite{chen2020automated}. In this paper, a new line of research is proposed that uses machine learning techniques to autonomously identify and separate desired wind events, thunderstorms in the present case, from large volumes of continuous data. Since the data from the wind monitoring networks are stored as time series, the use of an efficient time series representation named “Shapelet transform” has been proposed in this paper that is combined with a machine-learning algorithm to identify thunderstorm from anemometric records. The Shapelet Transform is a unique time series representation technique that is solely based on the time series shape \cite{lines2012shapelet}. For example, in terms of wind speed, every set of extreme wind event have their own unique time series signature as they pass over any instrumented station. The Shapelet Transform technique easily captures these unique time series shapes corresponding to each extreme event and the machine learning algorithm uses these shapes to identify thunderstorms from a large database. Moreover, a time series shape-based approach can unravel unseen corners of random phenomena like wind, which might have been overlooked by analysts using conventional meteorological and wind engineering parameters.

Section 2 gives a general overview of the shapelet transform and elaborates on the five major stages in the transform with the help of illustrative examples. Section 3 describes the wind monitoring network that has been used in this study. The network has been installed in the ports of the High Tyrrhenian Sea for the “Wind and Ports” (WP) and “Wind, Ports and Sea” (WPS) projects funded by the European Cross-border program to investigate extreme wind events in port areas. Section 4 explains the preprocessing of the raw wind velocity data acquired from the monitoring network. In particular, a wavelet-based denoising method is employed to remove a significant amount of noise while retaining the important features in the signal even when the noise is non-uniform. In section 5, the different shapes discovered from the wind speed measurements are elaborated. A comprehensive summary of the new thunderstorms detected using these shapes has been explained in great detail. Section 6 provides a synopsis of the efficient shapelet transform-based thunderstorm detection method.

\section{OVERVIEW OF SHAPELET TRANSFORM}

 Wind speed measurements recorded by an anemometer during a thunderstorm and during normal periods is shown in Fig.1. A sharp peak in wind velocity appears for a short duration in the thunderstorm time series that differs significantly from the other time series. These local shapes have very high ability to differentiate and can be used to distinguish between various types of time series. Thus, local shapes that may occur anywhere in a time series with high discriminatory power are termed as shapelets. In the present case, wind velocity shapelets serve as a dominant attribute for identifying thunderstorms from a large anemometric database containing continuous records. The identified shapelets are used to transform the original wind speed data where each attribute of the transformed dataset denotes the distance between a time series and a shapelet \cite{lines2012shapelet}. The transformed dataset is then applied to a machine learning algorithm, that identifies thunderstorms.
 \begin{figure}[htbp]
  \centering
  \captionsetup{justification=centering}
  \includegraphics[scale=0.8]{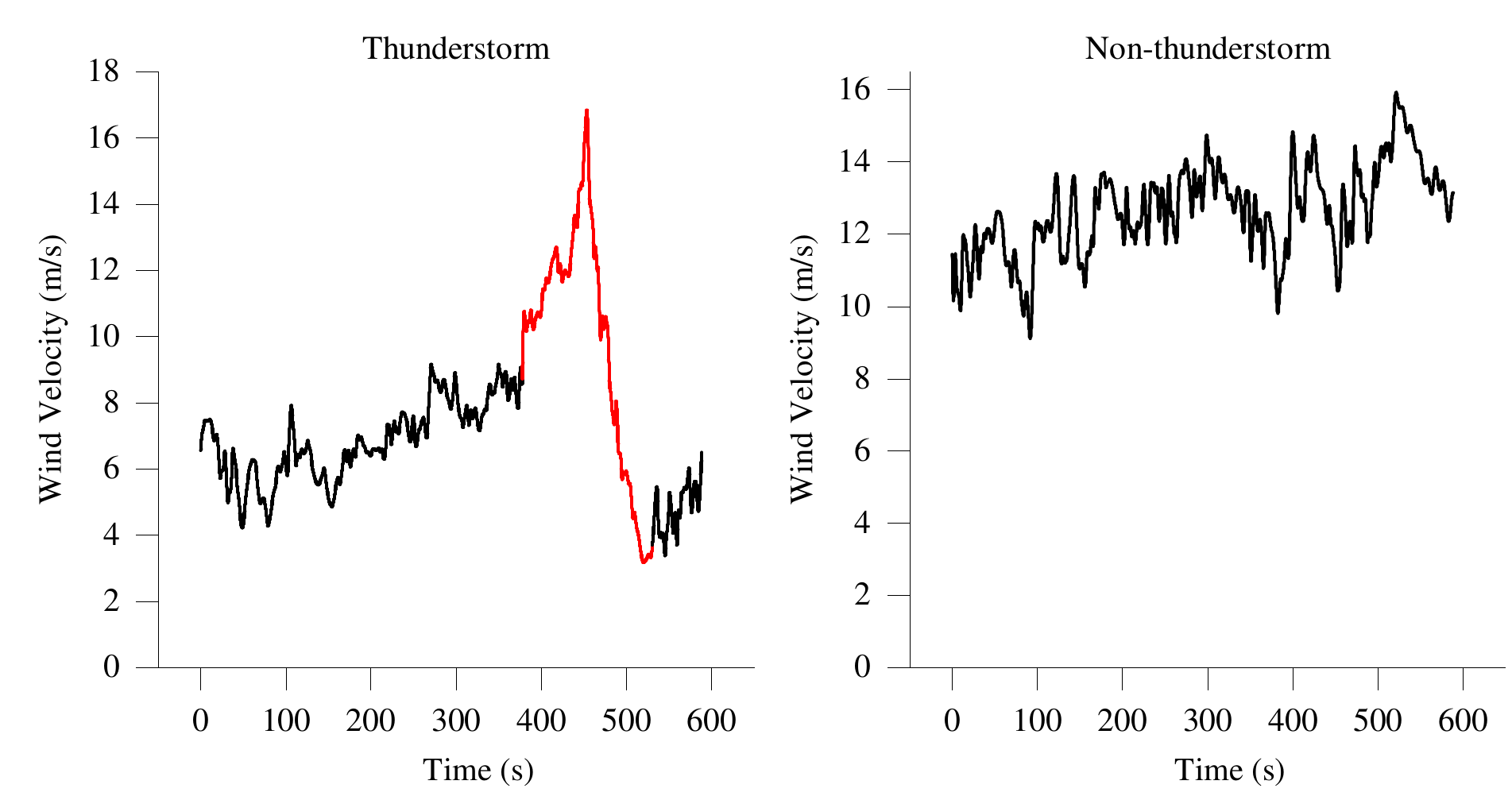}
  \caption{Time series of wind speed measurements for a thunderstorm (shapelet in red) vs a non-thunderstorm}
  \label{fig:fig1}
\end{figure}

\section{WIND MONITORING NETWORK AND DATASET}

\begin{figure}[htbp]
  \centering
  \captionsetup{justification=centering}
  \includegraphics[scale=1.3]{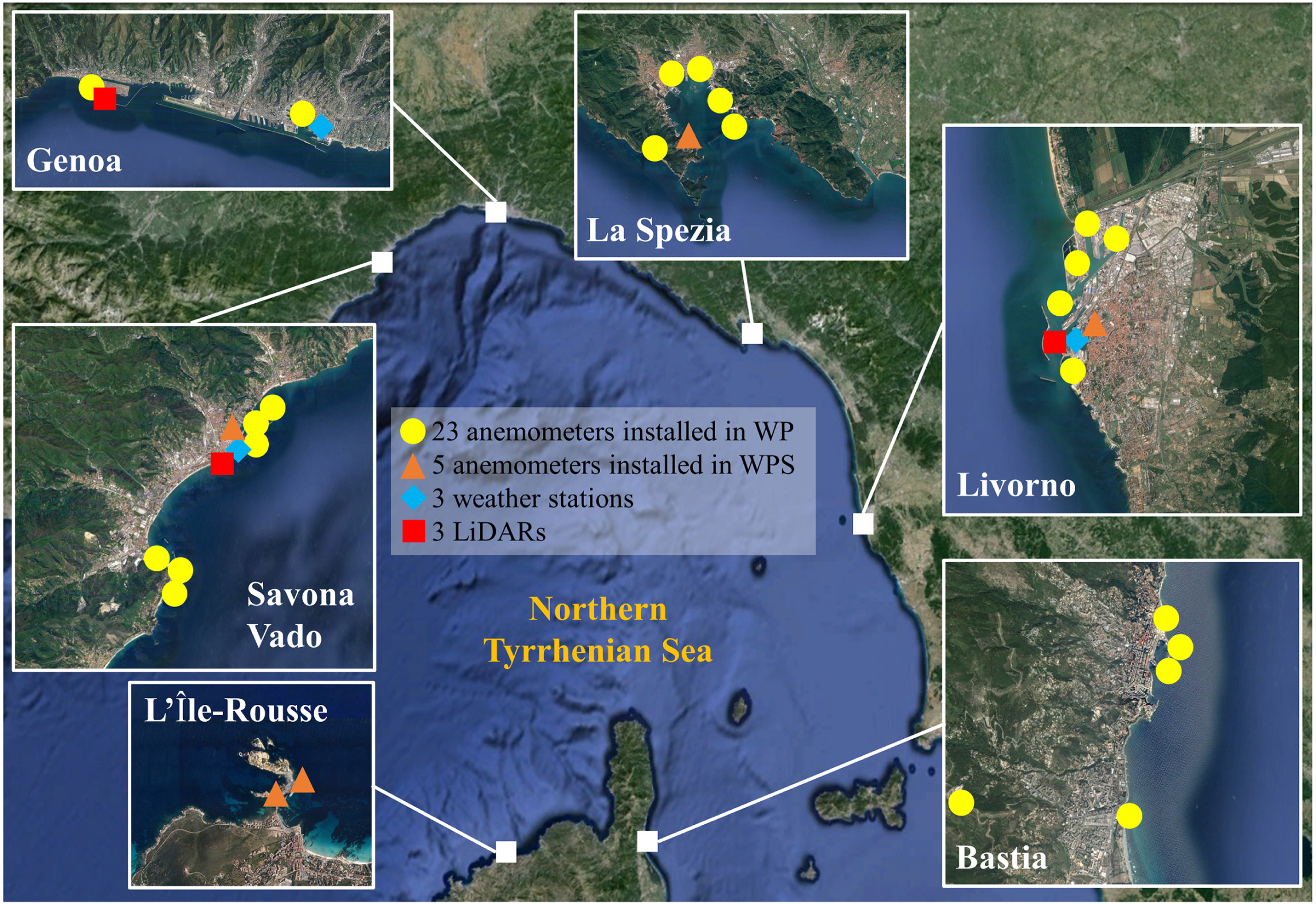}
  \caption{Anemometric stations at the ports that are part of the “Wind and Ports” and “Wind, Ports and Sea” project}
  \label{fig:fig2}
\end{figure}

Fig.2 shows the main features of the “Wind and Ports” (WP) \cite{solari2012wind} and “Wind, Ports and Sea” (WPS) \cite{repetto2017integrated,repetto2018web} projects financed by the European Cross-border program “Italy–France Maritime 2007-2013”. These projects are involved in the safe wind management and risk assessment for selected ports in Italy and France with the help of an extensive in-situ wind monitoring network.WP consists of 23 ultrasonic anemometers distributed in the ports of Genoa, La Spezia, Livorno, Savona – Vado Ligure, and Bastia. WPS, an enhancement of the WP network, consists of five additional ultrasonic anemometers installed in the ports of Savona, LaSpezia, Livorno, and L’Île-Rousse. The ultrasonic anemometers measure wind speed and direction with a precision of 0.01 m/s and 1° respectively. The sampling rate of anemometers is set to 10 Hz for the Ports of Genoa, La Spezia, and Livorno. One anemometer in the Port of Savona has a sampling frequency of 1 Hz while the others are set to 10 Hz. The anemometers in the Ports of Bastia and L’Île-Rousse have a sampling frequency of 2 Hz. Apart from anemometers, three Lidar (Light Detection and Ranging) wind profilers and three weather stations, each comprising of an additional ultra-sonic anemometer, thermometer, barometer, and a hygrometer are also installed in the ports of Genoa, Savona, and Livorno. A detailed description of the 
installed Lidars can be found in \cite{repetto2018web}. 

The instruments are positioned uniformly across the port areas and are installed on high rise towers or antenna masts atop buildings at least 10 m above the ground level to record undisturbed wind speed measurements. Local servers located in the head office of each of these ports receive data from the instruments and buffers accumulate the data over pre-defined intervals (10-min) for analyzing basic statistics such as average, peak wind speed and mean wind direction. The servers then send the raw data and statistics to a central server located at the Department of Civil, Chemical and Environmental Engineering (DICCA), the University of Genoa where the data is systematically checked and validated before storing it in a database.

\cite{zhang2018refined} used the semi-automatic, gust factor-based thunderstorm detection approach developed in \cite{de2014separation} to identify 10-min, 1-hour and 10-hour thunderstorm records from 14 ultrasonic anemometers. . The thunderstorms are referred to as “10-min” and “1-h” depending on whether the presence of a ramp-up and a transient peak is clearly detectable over a 10-min or 1-h long record. These thunderstorms have been used as ground truth for the present study and are referred to here as “cataloged thunderstorms”. The analysis carried out in this paper is limited to the data gathered by the 14 ultrasonic anemometers used in \cite{zhang2018refined}.  Table 1 shows the main properties of the anemometers used in this study along with their periods of measurement.

\begin{table}
\centering
\captionof{table}{Details of the anemometers used in the study}
\begin{tabular}{cccccc} 
\hline
Port                            & \begin{tabular}[c]{@{}c@{}}Anemometer \\ Type\end{tabular} & \begin{tabular}[c]{@{}c@{}}Anemometer \\ Number\end{tabular} & \begin{tabular}[c]{@{}c@{}}Height above \\ ground level (m)\end{tabular} & \begin{tabular}[c]{@{}c@{}}Period of \\ measurement\end{tabular} & \begin{tabular}[c]{@{}c@{}}Sampling \\ frequency (Hz)\end{tabular}  \\ 
\hline
\multirow{2}{*}{Genoa (GE)}     & \multirow{2}{*}{Biaxial}                                   & 1                                                            & 61.4                                                                     & 2011 – 2013                                                      & \multirow{2}{*}{10}                                                 \\
                                &                                                            & 2                                                            & 13.3                                                                     & 2010 – 2015                                                      &                                                                     \\
\multirow{2}{*}{La Spezia (SP)} & \multirow{2}{*}{Biaxial}                                   & 2                                                            & 13                                                                       & 2010 – 2015                                                      & \multirow{2}{*}{10}                                                 \\
                                &                                                            & 3                                                            & 10                                                                       & 2011 – 2015                                                      &                                                                     \\
\multirow{5}{*}{Livorno (LI)}   & \multirow{5}{*}{Triaxial}                                  & 1                                                            & 20                                                                       & \multirow{2}{*}{2010 – 2015}                                     & \multirow{5}{*}{10}                                                 \\
                                &                                                            & 2                                                            & 20                                                                       &                                                                  &                                                                     \\
                                &                                                            & 3                                                            & 20                                                                       & 2010 – 2015                                                      &                                                                     \\
                                &                                                            & 4                                                            & 20                                                                       & 2010   – 2015                                                    &                                                                     \\
                                &                                                            & 5                                                            & 75                                                                       & 2010 – 2014                                                      &                                                                     \\
\multirow{5}{*}{Savona (SV)}    & \multirow{5}{*}{Triaxial}                                  & 1                                                            & 33.2                                                                     & \multirow{5}{*}{2011   – 2015}                                   & \multirow{5}{*}{10}                                                 \\
                                &                                                            & 2                                                            & 12.5                                                                     &                                                                  &                                                                     \\
                                &                                                            & 3                                                            & 28                                                                       &                                                                  &                                                                     \\
                                &                                                            & 4                                                            & 32.7                                                                     &                                                                  &                                                                     \\
                                &                                                            & 5                                                            & 44.6                                                                     &                                                                  &                                                                     \\
\hline
\end{tabular}
\end{table}

\section{PREPROCESSING OF WIND SPEED MEASUREMENTS}

As mentioned in the previous section, the thunderstorms analyzed in this study have varying duration. Some of the thunderstorms are detectable over a 10-min period, while others are clearly detectable only over a 1-hour period. So, in the present study, a fixed window size of 1-hour is used to break down the continuous wind speed measurements that are sampled at 10 Hz. This window size is chosen as it can be used to identify both short and long duration thunderstorms. Moreover, thunderstorms (isolated or not) have a wind-producing duration of no more than one hour. Even if the sampling segment contains only a short part of the thunderstorm event, like the ramp up or ramp down segment, the ML-based algorithm proposed in this paper will still be able to identify and label these thunderstorms. More information on this is provided in section 6.

Close inspection of the data revealed several periods of measurement in which the recordings were extremely noisy and unreliable. It is thought that the majority of these noisy measurements were a result of interference from the radar on the approaching ships in the port areas. Excluding these measurements during the analysis will lead to a significant loss of valuable data. Hence, it is imperative to remove noise from the wind speed measurements to restore the completeness of the database. The use of canonical filters holds little utility in this case as the noise characteristics are unknown. Moreover, great care needs to be taken during the denoising process so that important features of the signal, such as spikes, are preserved as these sharp features can aid the shapelet transform in identifying the occurrence of thunderstorms. For this purpose, a robust denoising procedure based on Stationary Wavelet Transform (SWT) \cite{nason1995stationary} is used in this study as wavelets localize features in the data to different scales and can retain signal features while removing noise. 

SWT has a better performance in terms of denoising when compared to the Discrete Wavelet Transform (DWT), as the former has the properties of shift and scale invariance which is absent in the latter. The SWT-based denoising procedure is as follows
\begin{itemize}
 \item Suitable mother wavelet and decomposition level are determined for denoising
 \item Wavelet Transform is applied to the signal to obtain the wavelet coefficients
 \item Wavelet transform concentrates the signal features in large-magnitude coefficients and the small-magnitude coefficients are typically noise. A suitable threshold method and an appropriate threshold limit are applied to each level to remove noise.
 \item The signal is then reconstructed by applying the inverse wavelet transform of the thresholded coefficients. 
\end{itemize}

Following the above-mentioned procedure, SWT is applied to the noisy signals using Daubechies db10 as the mother wavelet with a maximum of 8 levels of decomposition. Soft fixed form thresholding is then applied to the wavelet coefficient to remove noise from the signal. Fig. 3 illustrates the effectiveness of this procedure. It can be seen from the figure that in both cases, SWT-based denoising has removed a significant amount of the noise while retaining the major features in the signal even when the noise is non-uniform. Once the measurements are denoised, a wind velocity threshold is applied: 1-s peak velocity over 1 hour > 10 m/s. All of the thunderstorms identified by \cite{zhang2018refined} and \cite{de2014separation} using the gust-factor approach have peak wind velocities above 15 m/s. Since the current study strives to find non-stationary records that were previously undiscoverable, a relatively low wind velocity threshold has been used. This way, a greater number of wind records can be analyzed using machine learning to search for the presence of thunderstorms. 

\begin{figure}[htbp]
  \centering
  \captionsetup{justification=centering}
  \includegraphics[scale=0.8]{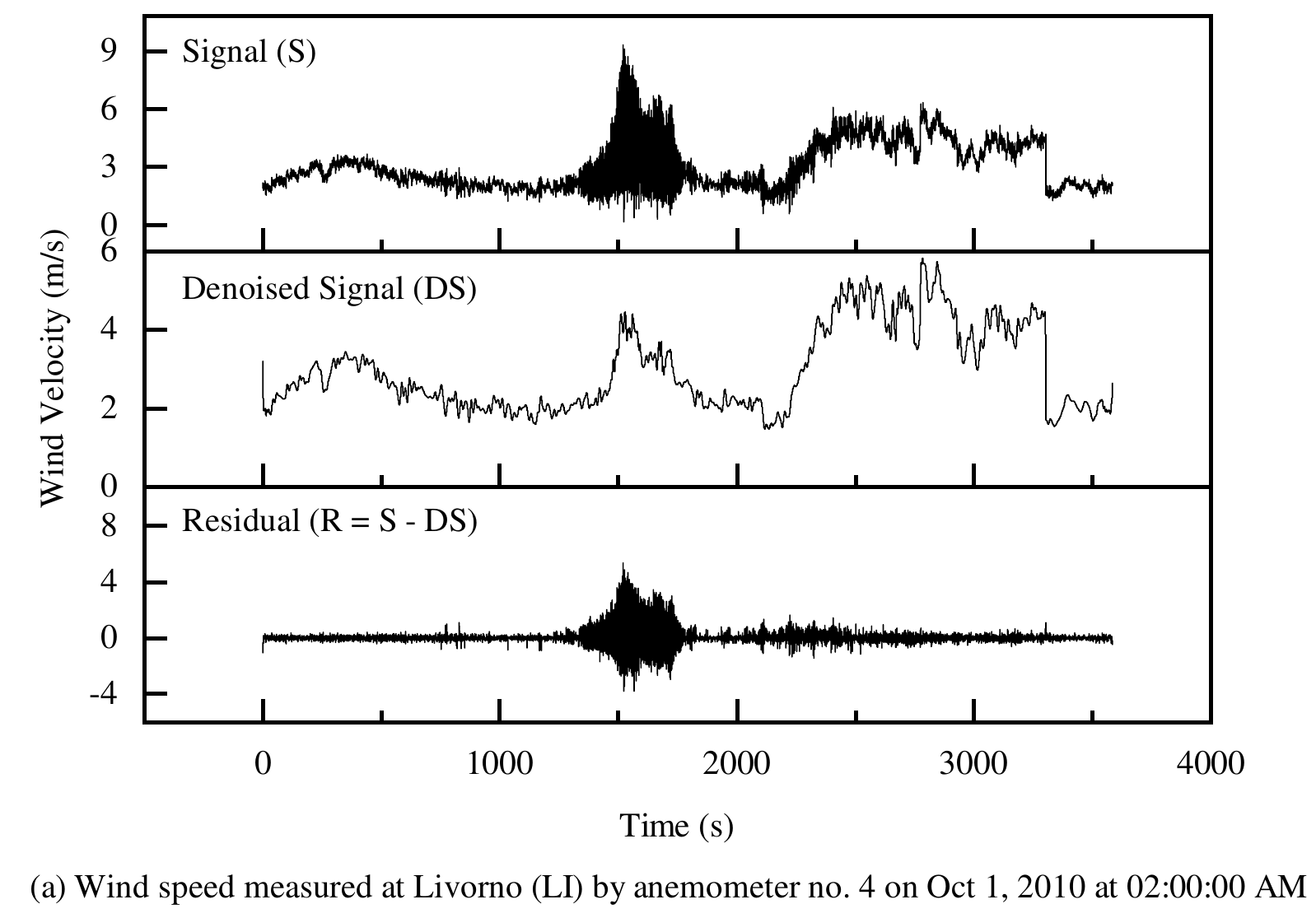}
    \includegraphics[scale=0.8]{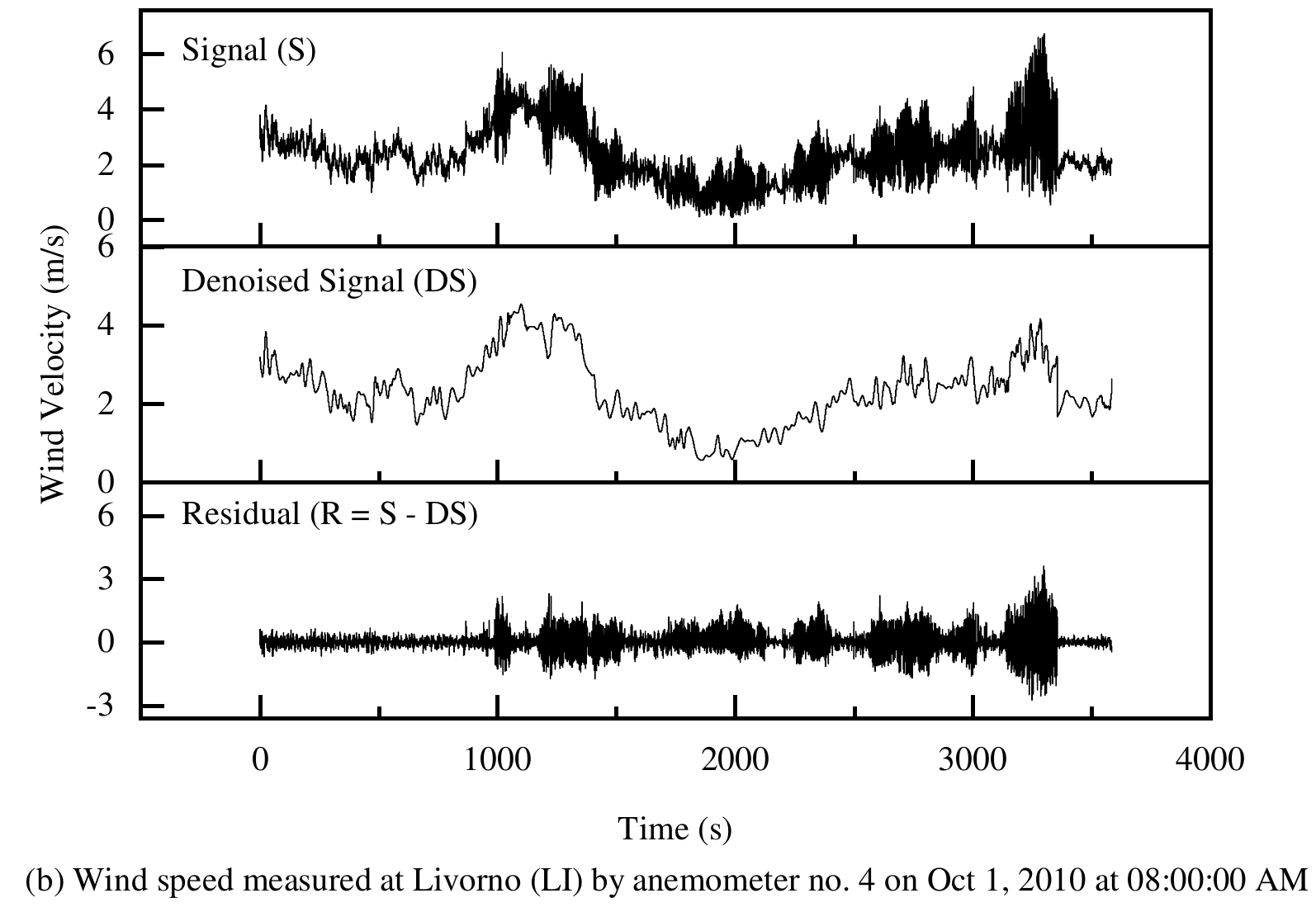}
  \caption{Wavelet-based denoising procedure to remove noise from the raw anemometric records }
  \label{fig:fig3}
\end{figure}

\section{METHODOLOGY FOR THUNDERSTORM IDENTIFICATION}

\begin{figure}[htbp]
  \centering
  \captionsetup{justification=centering}
  \includegraphics[width=\textwidth,scale=1.0]{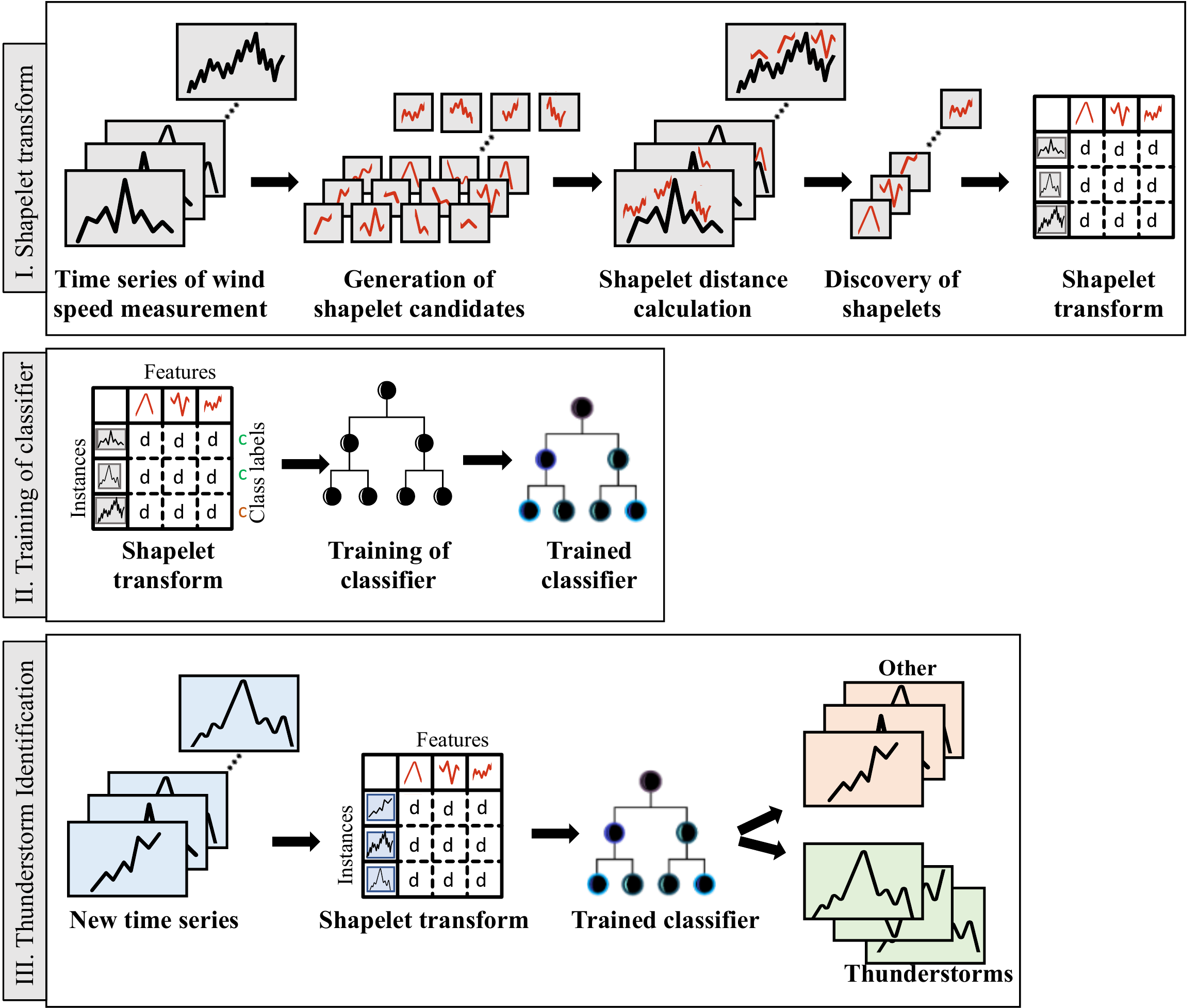}
  \caption{Methodology for thunderstorm identification using shapelet transform}
  \label{fig:fig4}
\end{figure}
The identification of thunderstorms involves three major stages as shown in Fig. 4. In the first stage, the preprocessed wind speed measurements are used to create a labeled time series learning set. The time series learning set is then utilized to transform the original data through five steps: generation of candidate shapelets, calculation of distance between a time series and a shapelet, assessment of shapelet quality, discovery of shapelets, and shapelet-based transformation of data. In a machine learning context, the features are the discovered shapelets and the instances are the time series. Thus, every element in the shapelet transform is the minimum Euclidean distance between each discovered shapelet and every time series in the learning set. In the second stage, the shapelet transform along with the associated class labels are used to train a random forest classifier to identify thunderstorms. The trained classifier is then employed to detect thunderstorms from new wind speed measurements from the wind monitoring system in the third stage. Each of these stages are explained in detail in the following sections.

\subsection{Shapelet transformation of dataset}

\subsubsection{Time series of wind speed measurements}

As mentioned in Section 3, the thunderstorms identified by \cite{zhang2018refined} using the semi-automated procedure developed by \cite{de2014separation} is used as the ground truth. This procedure establishes a wind velocity threshold and uses gust factor to separate the dataset into cyclones, thunderstorms, and intermediate events and the expert judgment involves the visual check of wind velocity records. Using this procedure, a total of 198 thunderstorm events and 277 strongly non-stationary records corresponding to these events were identified from the 14 anemometers listed in Table 1. Of these, 120 records were detected by anemometers during 2010-2014 in the Port of Livorno, which is well known to experience frequent thunderstorms.  These 120 wind speed records along with other non-thunderstorm records are used to train the shapelet algorithm to identify thunderstorms. The trained algorithm is then used to detect thunderstorms from all anemometers in the year 2015 (no measurements are available in 2015 for anemometer no. 1 in Genoa, so the year 2012 is considered) and the results are compared with the cataloged events. 

\begin{figure}[htbp]
  \centering
  \captionsetup{justification=centering}
  \includegraphics[scale=0.7]{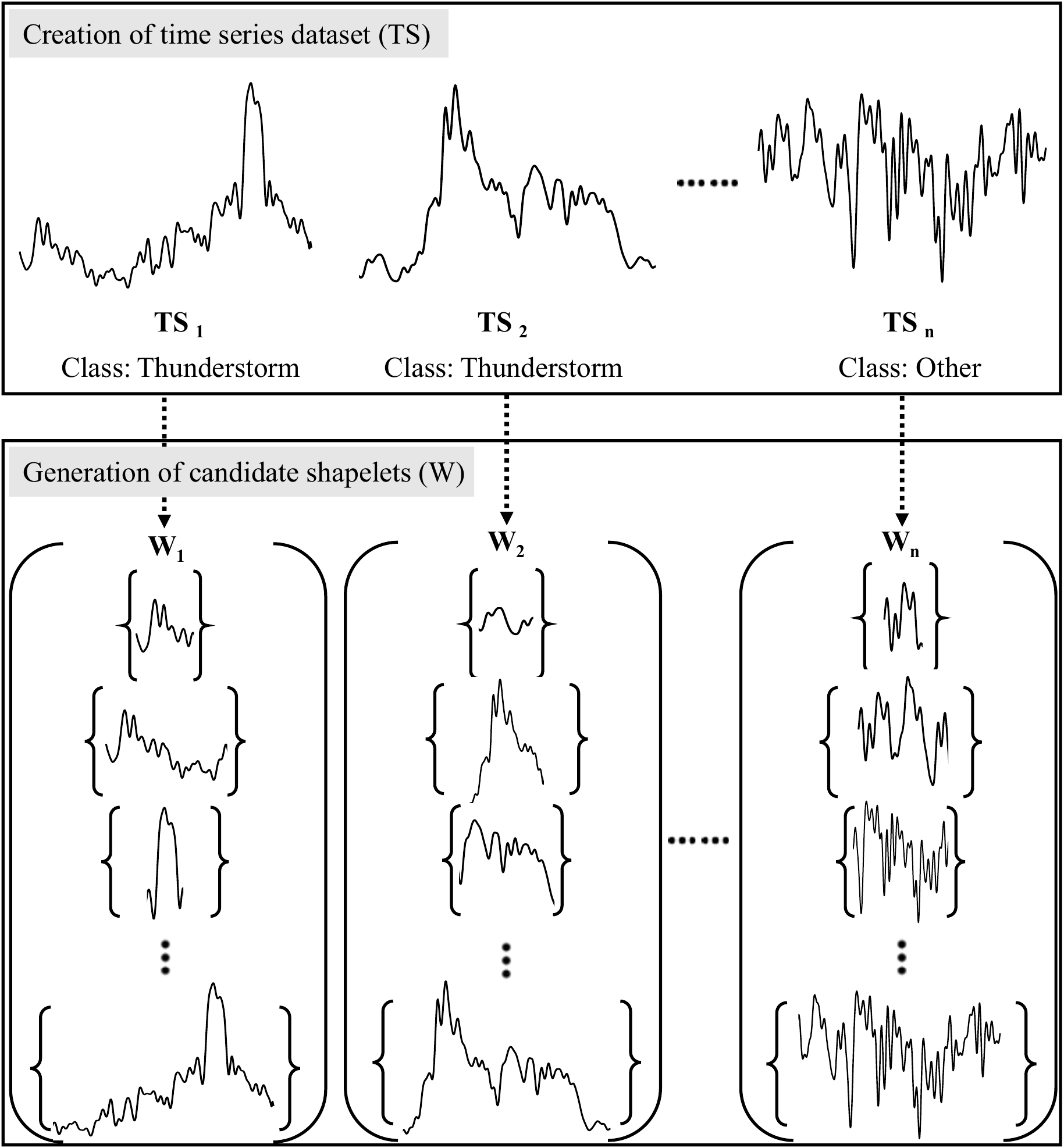}
  \caption{Illustration of candidate shapelet generation for each time series}
  \label{fig:fig5}
\end{figure}

\subsubsection{Generation of shapelet candidates}

 Let $TS=\left\{T S_{1}, T S_{2} \ldots \ldots, T S_{n}\right\}$ be a time series dataset where $T S_{i}=\left\langle t_{i, 1}, t_{i, 2}, \ldots, t_{i, m}\right\rangle$ is an individual time series containing wind speed measurements. Let the class labels for each time series be denoted by C. Here the class labels are “Thunderstorm” and “Other”. A set of input-output pairs is created using the individual time series and their associated labels which serve as the learning set, $\Phi=(T S, C)$.  In the present case, the learning set for the algorithm consists of 120 thunderstorm records labeled as “Thunderstorms” and 120 non-thunderstorm records labeled as “Other”. The learning set contains equal samples from both the classes to avoid any classifier bias during the identification of thunderstorms. The dataset is then randomly split into training (70\%) and test (30\%) sets. So, the training set contains 168 time-series records, and the test set contains 72 records. The time series in training set is used in the following steps to discover shapelets and the efficacy of the method is tested on the time series in the testing set.
 
 As shown in Fig.5, each subsequence of a time series is regarded as a prospective shapelet candidate. The smallest subsequence contains three data points as it is the shortest possible length for a time series and the largest subsequence is the length of the time series. Thus, if $m$ denotes the length of an individual time series and $l$ denotes the length of a subsequence, then there are $(m-l)+1$ distinct subsequences in any individual time series. In the present study, the continuous wind speed measurements are broken down into 1-hour intervals sampled at 10 Hz. Thus, each time series in the training set has 36000 data points. Let us take the first time series in the training set for illustration. The set of all candidate shapelets for this time series is
\begin{equation}
	W_{1}=\left\{w_{3}, w_{4}, \ldots ., w_{35988}, w_{35999}, w_{36000}\right\}
\end{equation}
where $w_{3}$ (first three data points) is the shortest shapelet length and $w_{36000}$ (entire time series) is the longest shapelet length. . It should be noted that the shapelet algorithm independently normalizes all candidate shapelets so as to be invariant to scale and offset. Thus, the set $W_{1}$ contains $w_{35998}$ different lengths of shapelets. In a similar way, shapelet candidates are generated from all of the time series in the learning set.

\subsubsection{Calculation of distance between a shapelet and a time series}

\begin{figure}[htbp]
  \centering
  \captionsetup{justification=centering}
  \includegraphics[scale=0.22]{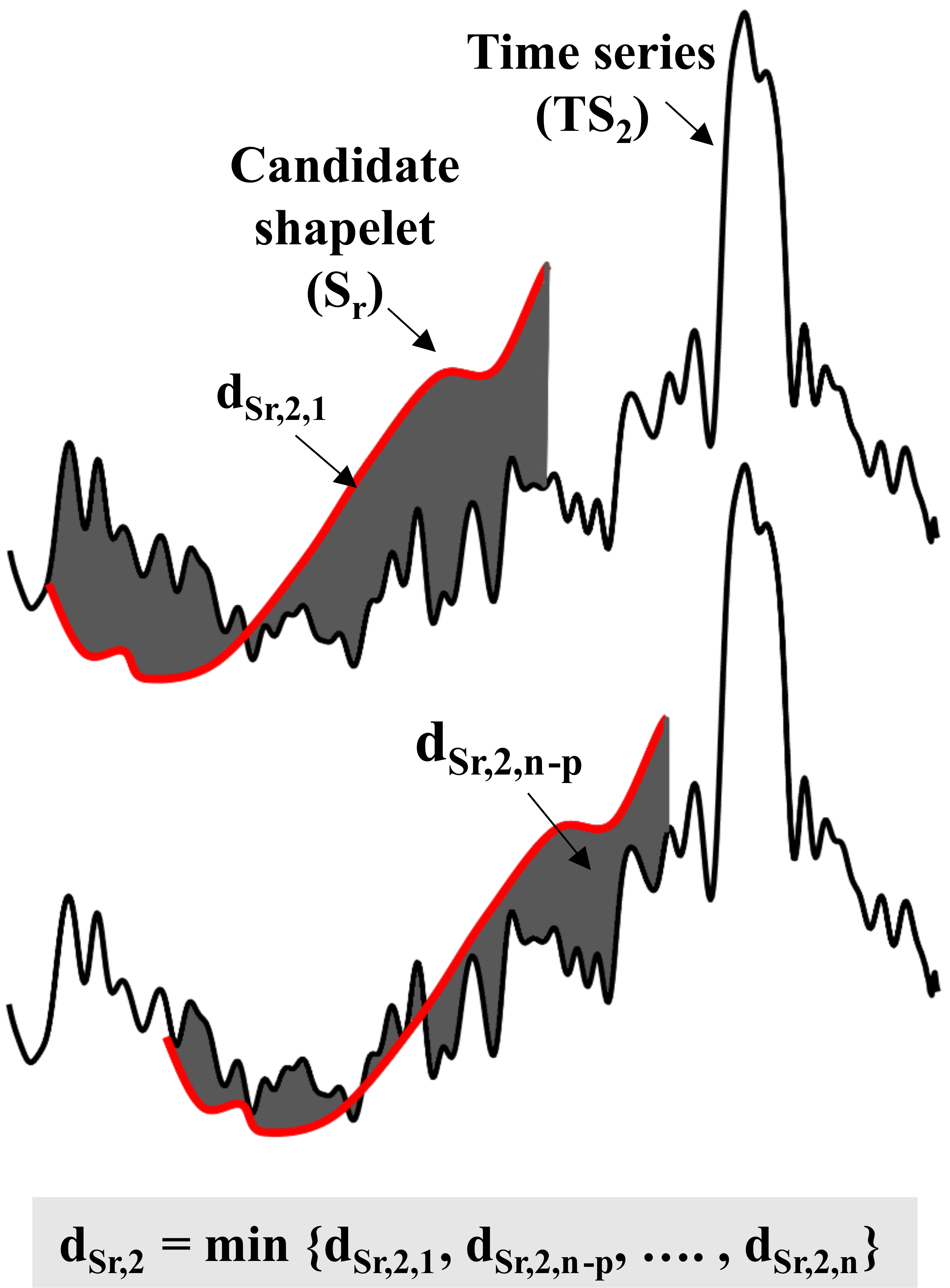}
  \caption{Illustration of calculation of Euclidean distance between a candidate shapelet and a time series }
  \label{fig:fig6}
\end{figure}

The similarity between a shapelet and a time series is measured using Euclidean distance. The Square Euclidean distance between two subsequences $X$ and $Y$ of the same length $l$ is $\sum_{i=1}^{l}\left(x_{i}-y_{i}\right)^{2}$. Consider a shapelet candidate $S_{r}$ and a time series $T S_{2}$ as shown in Fig.6. The distance between $S_{r}$ and $T S_{2}$ is the minimum of all the Euclidean distances calculated. If the time series contains a shape very similar to the candidate shapelet, the Euclidean distance will be very low and vice versa. Once the distance between a shapelet candidate and each and every time series in $TS$ is calculated this way, an orderline $DS$ is created that contains the list of these distances along with their class label. The orderline is sorted in ascending order of the distance value. Thus, if there are $n$ time series, the orderline ($D_{S_{r}}$) for a shapelet candidate $S_{r}$ is given by $D_{S_{r}}=\left\langle d_{S_{r, 1}}, d_{S_{, 2}, \ldots, \ldots} d_{S, m}\right\rangle$.
	
In the present study, each time series leads to the generation of 35998 shapelet candidates as seen in Eq. (1). Each of these 35998 shapelets is compared with other time series using Euclidean distance. So, for the present case, $W_{3}$ is compared with the 167 other time series in the training set using a minimum Euclidean distance. The thus obtained 167 distances values are ordered in increasing value to create the orderline. Then $W_{4}$ is compared with the 167 other time series and so on.

\subsubsection{Assessment of shapelet quality}

\begin{figure}[htbp]
  \centering
  \captionsetup{justification=centering}
  \includegraphics[scale=0.7]{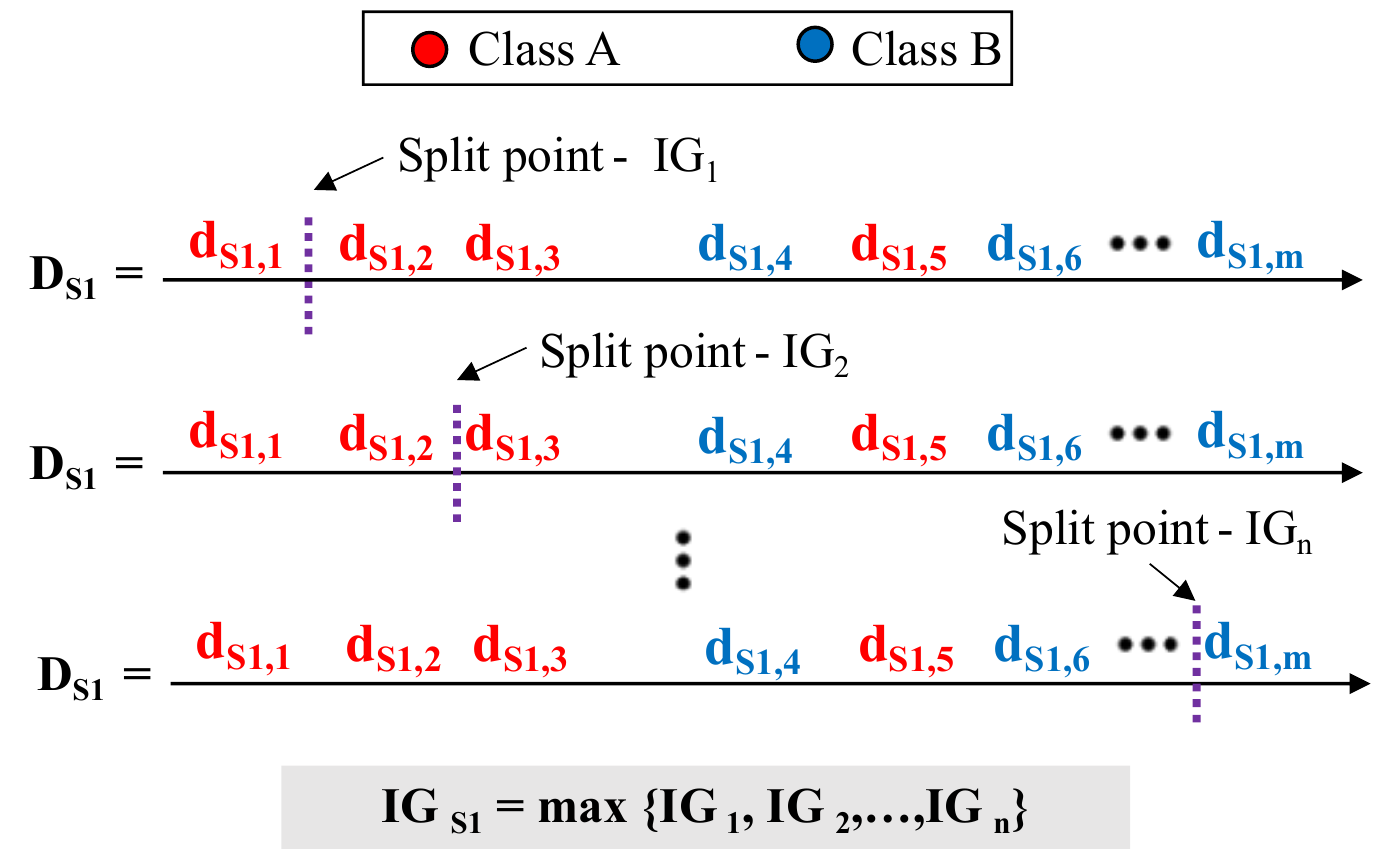}
  \caption{One-dimensional representation of the arrangement of time series objects by the distance to the candidate shapelet S1. Information Gain is calculated for each possible split point on the orderline DS1}
  \label{fig:fig7}
\end{figure}

\begin{figure}[htbp]
  \centering
  \captionsetup{justification=centering}
  \includegraphics[scale=0.9]{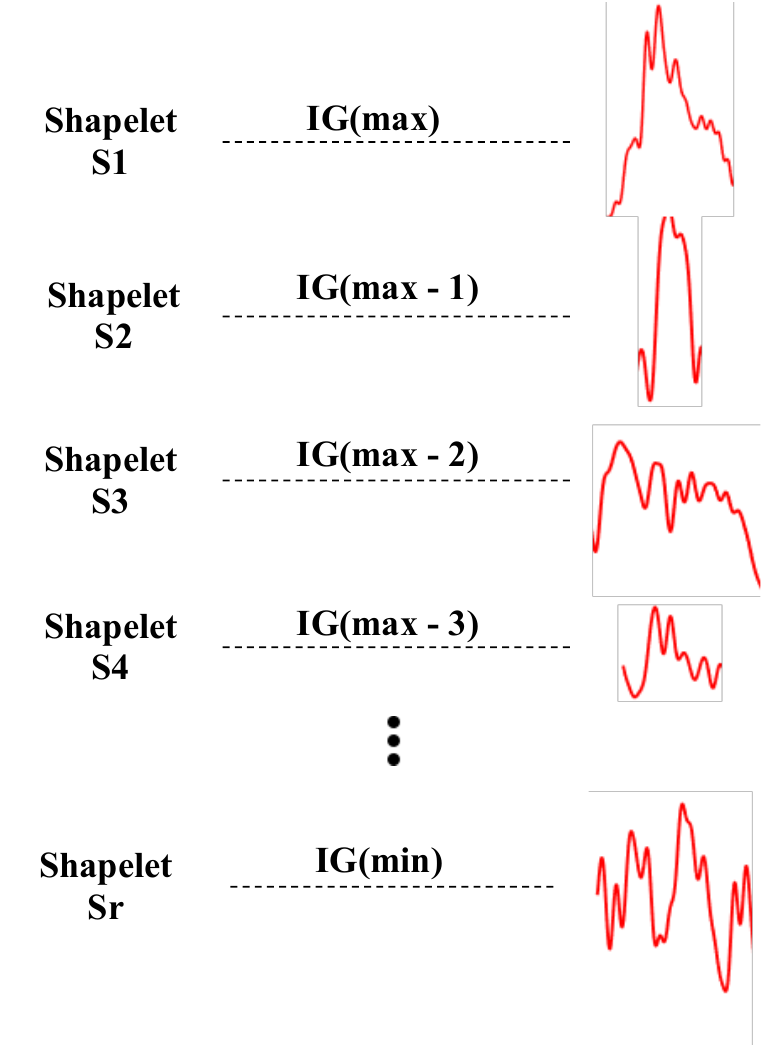}
  \caption{Discovery of shapelets based on Information Gain}
  \label{fig:fig8}
\end{figure}

A large number of shapelet candidates are generated from the previous step. It is not computationally efficient to store all the shapes irrespective of their quality. To optimize the process, Information Gain \cite{shannon1949mathematical} is used for testing the quality of the various captured shapes \cite{mueen2011logical,ye2009time,ye2011time}. To illustrate the concept of information gain, consider a dataset $S$ which has two classes $A$ and $B$. The randomness in the dataset is measured in terms of entropy. The entropy of $S$ is given by: 

\begin{equation}
    E(S)=-p(A) \log _{2}(p(A))-p(B) \log _{2}(p(B))
\end{equation}
	
where $p(A)$ and $p(B)$ are the number of objects in each of these classes. Entropy takes a value between 0 and 1. A high entropy suggests a low level of purity among the classes and most of the machine learning algorithms aim to reduce the entropy. A metric that is used to measure the reduction in entropy after a dataset is split based on an attribute is called Information Gain (IG). Consider an attribute that splits the dataset $T$ into two datasets $T_{A}$ and $T_{B}$. The IG of this split is given by

\begin{equation}
   I G=E(S)-\left(\frac{\left|T_{A}\right|}{|T|} E\left(T_{A}\right)+\frac{\left|T_{B}\right|}{|T|} E\left(T_{B}\right)\right)
\end{equation}

where $0\le IG\le 1$.
A high information gain denotes the high informative power of an attribute. This way the least informative attributes can be abandoned. Here, shapelets are the attributes and the orderline ($DS$) containing distances between the shapelet candidate and the time series is split in various ways and the IG of each of these splits are compared as shown in Fig. 7. An optimal shapelet generates large distance values for a time series that does not belong to its own class and small distance values otherwise. A best split in the orderline has all the distance values of a particular class on one side of the split and the rest of the distance values on the other side. This split produces the highest IG.  This way, the highest IG obtained by each of the shapelet candidates are calculated. The shapelets are then arranged in a decreasing order based on IG as shown in Fig. 8. Whichever shapelet surpasses the minimum provided information gain threshold (0.05 in the present case) is retained and the other shapelets are discarded. This makes sure that the selected shapelets are meaningful and have discriminatory power. 
\vspace{\baselineskip}

\subsubsection{Discovery of shapelets}

The shapelet algorithm was developed by \cite{bagnall2017great} and the full workflow along with the code is available at \url{www.sktime.org} \cite{loning2019sktime}. The same algorithm has been used but with slight modifications to suit the datasets in the present study. Detailed information about the application of the algorithm is also available in \cite{arul2020applications}. The algorithm is straightforward, and no parameter tuning is required. The only requirement for the shapelet algorithm is the input time series ($TS$). For the present case, a total of 32 shapelets are discovered from the training set using the shapelet algorithm. The first six (highlighted in red) shapelets with the highest IG among the 32 are shown in Fig.9. From shapelets 1-2 and 4-6, it can be seen that the peaks in the time series are extracted as shapelets. A section of time series from a non-thunderstorm wind speed measurement is also captured as a shapelet as seen in Shapelets 3. As the current study involves two classes (thunderstorm and non-thunderstorm), two groups of shapelets are discovered, one for each class. Of the 32 shapelets, 11 correspond to non-thunderstorm records. This may raise another question as to why two families of shapelets are required for a binary classification problem instead of just using one set of shapelets for classification. Using only one family of shapelets on datasets generated due to erratic natural phenomena like the wind will affect the detection accuracy as these datasets contain many instances of time series where a clear A vs B cannot be established. Such time series can only be correctly identified if they are compared with both sets of shapelets.

\begin{figure}[htbp]
  \centering
  \captionsetup{justification=centering}
  \includegraphics[scale=0.8]{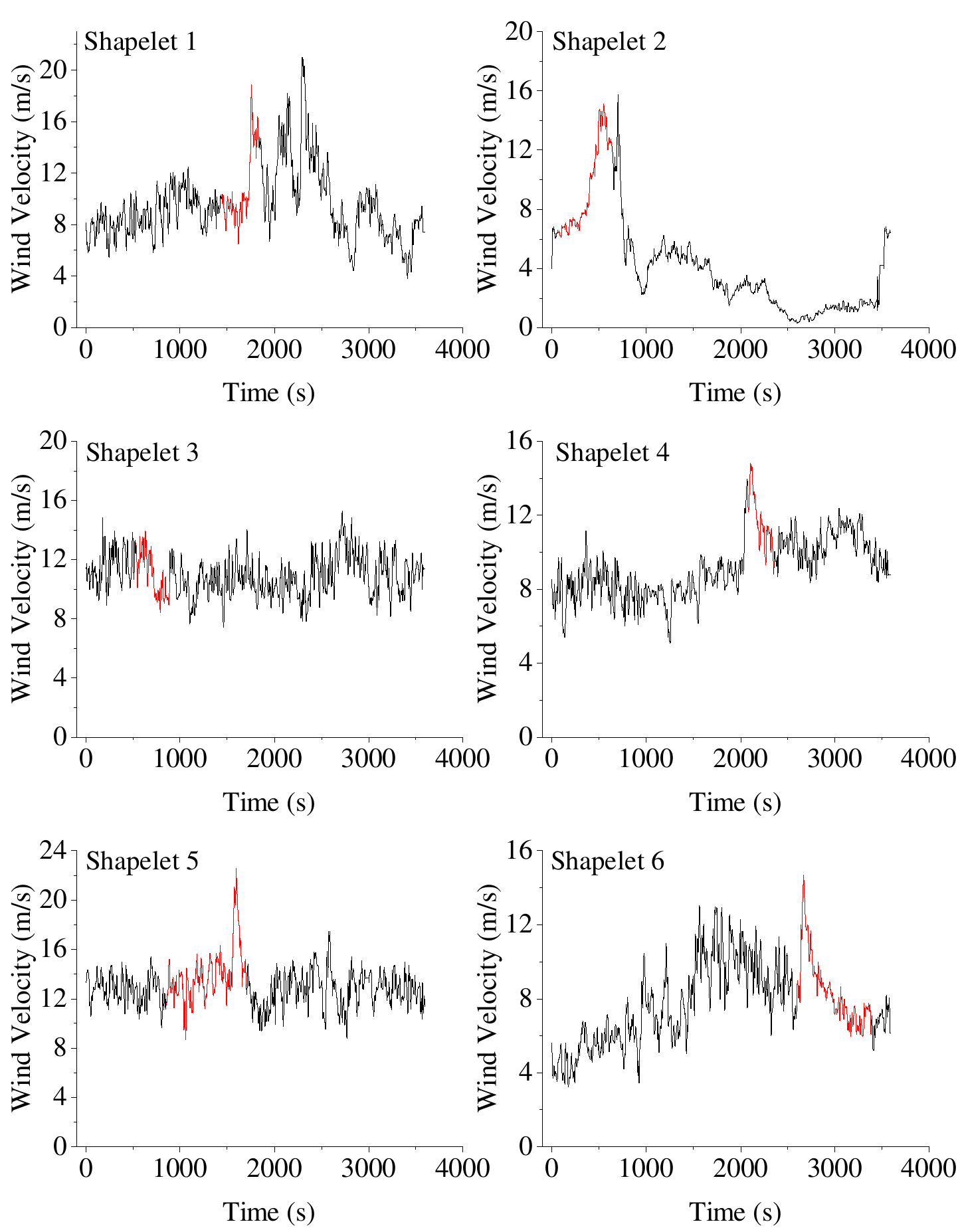}
  \caption{First six shapelets (highlighted in red) discovered for the identification of thunderstorms }
  \label{fig:fig9}
\end{figure}

It should be noted that, the discovered shapelets are of various lengths. Predetermining the optimal length of shapelet is impossible and unnecessary as it hinders the detection accuracy of the algorithm. It is also very difficult to interpret the variety of shapelet lengths obtained from the algorithm as these lengths have been chosen from several 1000s of shapelet lengths that were compared with several other time series. However, there is a provision in the shapelet algorithm to set the maximum and minimum shapelet length to achieve speedup. This provision should be used with care and should only be utilized in cases where only a certain length of shapelets are of interest.

\subsubsection{Shapelet transform}

Shapelet transform converts time series data into a local-shape space where each attribute denotes the distance between a time series and a shapelet \cite{hills2014classification, lines2012shapelet}. To train the machine learning algorithm for thunderstorm identification, the original time series in the training set ($\mathrm{T}_{\mathrm{R}}$$S$) is transformed to a local shape space using the 32 discovered shapelets. Thus a 168 x 32 matrix is constructed where each element is the minimum Euclidean distance between a time series and a shapelet in the training set. In a similar way, the testing set ($\mathrm{T}_{\mathrm{T}}$$S$) is also transformed using the discovered shapelets. In the present case, the testing set has 72 time series, and  a 72 x 32 matrix is constructed. In the context of machine learning, the shapelets are the features and the individual time series are the instances and the class labels are attached at the end of each instance as shown in Fig. 10. 
 \begin{figure}[htbp]
  \centering
  \captionsetup{justification=centering}
  \includegraphics[scale=0.9]{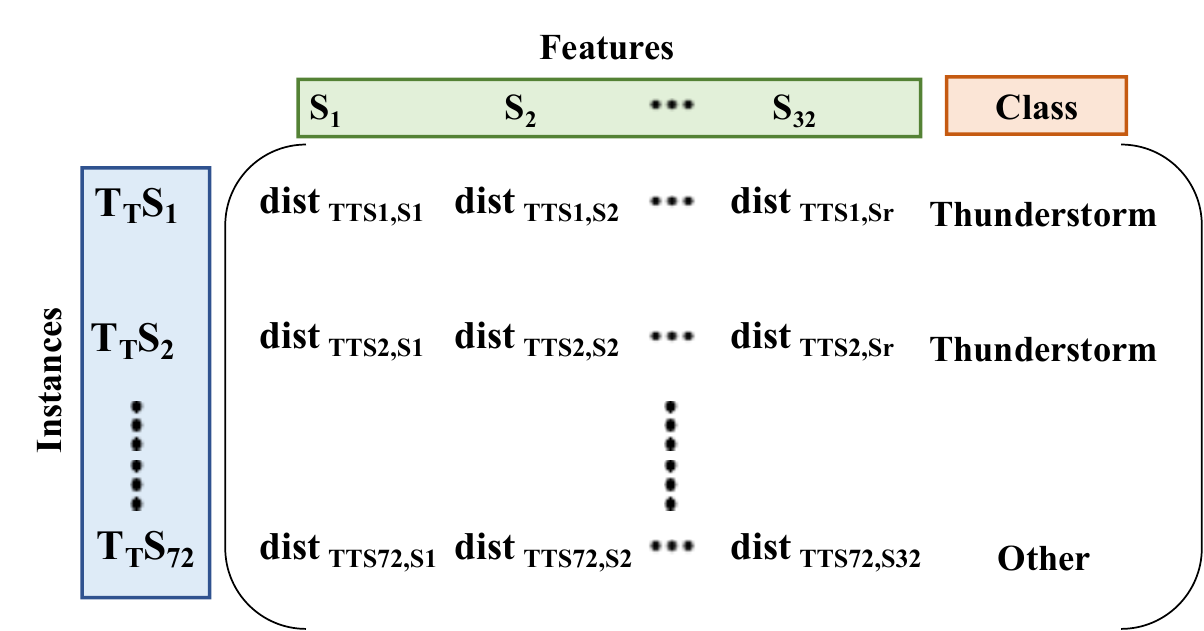}
  \caption{Shapelet Transform containing a matrix of Euclidean distance between the discovered shapelets and the time series in testing set}
  \label{fig:fig10}
\end{figure}

\subsection{Training of shapelet-based classifier}

Random Forest (RF) classifier \cite{breiman2001random} with 500 trees is used for the current study to identify thunderstorms from the shapelet transformed testing set. \cite{hills2014classification,bagnall2017great} compared the performance of shapelets using several standard classifiers and ensemble classifiers on a variety of datasets from UCR time-series repository. According to their study, a random forest classifier with 500 trees is found to be optimal on a shapelet-transformed dataset. Hence the same has been adopted in the present study. It is also found that increasing the number of trees beyond 500 did not result in any significant increase in accuracy. Moreover, along with each prediction, the classifier also gives a class probability. For instance, if a time series is predicted as Thunderstorm, the classifier also gives the prediction probability, i.e., prob (Thunderstorm) = 77\% and prob (Other) = 23\%. More insights about this is provided in the following sections.

\subsection{Identification of thunderstorms}

The Random Forest classifier is first trained on the shapelet transformed training dataset. No threshold is set for the depth of trees in the Random Forest. The nodes are allowed to expand until all the leaves are pure (that is, all the samples at that node have the same label). The average depth of the trained RF is 3 (obtained by taking the mean of the individual tree depth of all 500 trees). The trained classifier is then used on ${T}_{T}{S}$ (72 x 32) to test the performance of the model. The detection results and performance metrics for the classifier are shown in Table 2. The following will help understand the performance metrics better.

\begin{itemize}
 \item True Negative (TN) - The actual label is Other, and the classifier predicted Other.
 \item False Positive (FP) - The actual label is Other, and the classifier predicted Thunderstorm.
 \item False Negative (FN) - The actual label is Thunderstorm, and the classifier predicted Other.
 \item True Positive (TP) - The actual label is Thunderstorm, and the classifier predicted Thunderstorm.
 \item Precision (Pr) – For each class, precision is the ratio of the number of correct predictions to the number of total predictions made.
 \item Recall (Re) – For each class, recall is the ratio of correctly predicted observations to all actual class observations. 
 \item F1 Score - The harmonic mean of precision and recall and is a combined measure of the two. The highest possible value of an F-score is 1.0, and the lowest possible value is 0.
 \item Matthews Correlation Coefficient (MCC) - A measure of the quality of binary classification. It describes how changing the value of one variable will affect the value of another and returns a value between -1 and 1, where 1.0 describes a perfect prediction, 0 denotes unable to return any valid information (no better than random prediction), and -1 describes complete inconsistency between prediction and observation.
 \item Accuracy - The sum of true positives and true negatives divided by the total number of instances.
\end{itemize}

\begin{table}[h]
\centering
\captionof{table}{Performance metrics for the shapelet-based random forest classifier}
\begin{tabular}{ccccccc} 
\hline
Class                                                              & \begin{tabular}[c]{@{}c@{}}Predicted:\\Other\end{tabular} & \begin{tabular}[c]{@{}c@{}}Predicted:\\Thunderstorm\end{tabular} & \begin{tabular}[c]{@{}c@{}}Class \\Precision\end{tabular}   & \begin{tabular}[c]{@{}c@{}}Class \\Recall\end{tabular}      & \begin{tabular}[c]{@{}c@{}}F1 \\Score\end{tabular}                   & \begin{tabular}[c]{@{}c@{}}Matthews \\Correlation \\Coefficient \\(MCC)\end{tabular}  \\ 
\hline
\begin{tabular}[c]{@{}c@{}}\\Actual:\\Other \\\end{tabular}        & 26(TN)                                                    & 1(FP)                                                            & \begin{tabular}[c]{@{}c@{}}TN/(TN+FN) \\= 0.90\end{tabular} & \begin{tabular}[c]{@{}c@{}}TN/(TN+FP) \\= 0.96\end{tabular} & \begin{tabular}[c]{@{}c@{}}2*(Pr*Re) / \\(Pr+Re) = 0.95\end{tabular} & \multirow{2}{*}{0.89}                                                                 \\
\begin{tabular}[c]{@{}c@{}}\\Actual:\\Thunderstorm \\\end{tabular} & 3(FN)                                                     & 42(TP)                                                           & \begin{tabular}[c]{@{}c@{}}TP/(TP+FP) \\= 0.97\end{tabular} & \begin{tabular}[c]{@{}c@{}}TP/(TP+FN) \\= 0.93\end{tabular} & \begin{tabular}[c]{@{}c@{}}2*(Pr*Re) / \\(Pr+Re) = 0.93\end{tabular} &                                                                                       \\
\begin{tabular}[c]{@{}c@{}}\\Accuracy \\\end{tabular}              & \multicolumn{6}{c}{(TP+TN)/(TP+TN+FP+FN) = 0.95}                                                                                                                                                                                                                                                                                                                                                                        \\
\hline
\end{tabular}
\end{table}

From Table 2, it can be seen that 97\% precision is obtained for class “Thunderstorm” with an overall accuracy of 95\%. The F1 score is also high for both classes. MCC would return a high score only if the prediction obtained good results in all of the four confusion matrix categories (TP, FN, TN, FP), proportionally both to the size of positive elements and the size of negative elements in the dataset. For the present case, a high MCC score of 0.89 is obtained. The model performs well on all the three metrics, i.e., accuracy, F1 Score, and MCC, thus demonstrating the efficacy of the method to identify thunderstorms. This trained classifier is then used on the wind speed measurements from all anemometers to identify thunderstorms.

\section{RESULTS OF BINARY CLASSIFICATION}

Continuous measurements (sampled at 10 Hz) for a period of 1-year, obtained from 14 anemometers, is used in this study to identify thunderstorms. As mentioned in Section 4, the continuous measurements are broken down into 1-hour segments. After removing periods of no measurement, a total of 12,725 1-hour segments of data are obtained. Of these, 240 datasets are used for training and testing the algorithm as explained in Section 5.1.2.  So, the Shapelet-based Random Forest classifier is used to detect thunderstorm records from 12,485 datasets. A summary of the detection performance of Shapelets in terms of several metrics is provided in Table 3. The shapelet-based classifier was able to discover a total of 240 strongly non-stationary wind speed records, as shown in Table 3, that can be traced back to thunderstorm-like outflows. A record is referred to herein as non-stationary when it exhibits statistical irregularity for a time period of 1 hour. Apart from the 60 non-stationary catalog records that were originally identified based on the gust factor method, an additional 180 records were identified using shapelets. All the detections were cross-checked with radar images to separate true positives from false positives. It is worth noting that a thunderstorm is a local storm invariably produced by a cumulonimbus cloud and always accompanied by lightning and thunder. Therefore, as the current study does not check the occurrence of lightning strikes for all the detected thunderstorms, it is possible that some events belong to the wider class of severe convective storms (which includes thunderstorms as well) that do not necessarily produce lightning \cite{bluestein2013severe}, but can still bring about strong gusts of wind.

\begin{table}[h]
\centering
\captionof{table}{Performance metrics for the Shapelet-based Random Forest classifier for all anemometers}
\begin{tabular}{cccccccccc} 
\hline
\multirow{2}{*}{Port}                                                          & \multirow{2}{*}{Year} & \multirow{2}{*}{Anemometer} & \multicolumn{3}{c}{TP}                                                                                                                          & \multirow{2}{*}{TN} & \multirow{2}{*}{FN} & \multirow{2}{*}{FP} & \multirow{2}{*}{Accuracy}  \\ 
\cline{4-6}
                                                                               &                       &                             & \begin{tabular}[c]{@{}c@{}}No. of \\ catalog detections\end{tabular} & \begin{tabular}[c]{@{}c@{}}No. of \\ new detections\end{tabular} & Total &                     &                     &                     &                            \\ 
\hline
\multirow{2}{*}{Genoa}                                                         & 2012                  & 1                           & 7/7                                                                  & 32                                                               & 39    & 1359                & 4                   & 362                 & 0.80                       \\
                                                                               & 2015                  & 2                           & 1/2                                                                  & 7                                                                & 8     & 894                 & 2                   & 49                  & 0.96                       \\
\multirow{4}{*}{Livorno}                                                       & \multirow{4}{*}{2015} & 1                           & 11/11                                                                & 20                                                               & 31    & 561                 & 3                   & 51                  & 0.93                       \\
                                                                               &                       & 2                           & 7/7                                                                  & 21                                                               & 28    & 678                 & 1                   & 53                  & 0.94                       \\
                                                                               &                       & 3                           & -                                                                    & -                                                                & -     & 36                  & -                   & 7                   & 0.85                       \\
                                                                               &                       & 4                           & 5/5                                                                  & 25                                                               & 30    & 488                 & 1                   & 109                 & 0.83                       \\
\multirow{5}{*}{\begin{tabular}[c]{@{}c@{}}Savona \\ and \\ Vado\end{tabular}} & \multirow{5}{*}{2015} & 1                           & 3/5                                                                  & 12                                                               & 15    & 1110                & 3                   & 141                 & 0.90                       \\
                                                                               &                       & 2                           & 3/3                                                                  & 7                                                                & 10    & 1135                & 1                   & 143                 & 0.90                       \\
                                                                               &                       & 3                           & 4/6                                                                  & 10                                                               & 14    & 1277                & 5                   & 238                 & 0.85                       \\
                                                                               &                       & 4                           & 7/7                                                                  & 10                                                               & 17    & 1495                & 3                   & 228                 & 0.88                       \\
                                                                               &                       & 5                           & 4/4                                                                  & 9                                                                & 13    & 1136                & 6                   & 113                 & 0.92                       \\
\multirow{2}{*}{La Spezia}                                                     & \multirow{2}{*}{2015} & 2                           & 3/4                                                                  & 16                                                               & 19    & 151                 & 4                   & 25                  & 0.86                       \\
                                                                               &                       & 3                           & 5/6                                                                  & 11                                                               & 16    & 286                 & 1                   & 86                  & 0.80                       \\
\multicolumn{3}{c}{Total}                                                                                                            & 60                                                                   & 180                                                              & 240   & 10606               & 34                  & 1605                &                            \\
\hline
\end{tabular}
\end{table}

Figs. 11-15 show a typical l-hour record of a strongly non-stationary event detected by the shapelet method. The existence of thunderstorms associated with these records is proven by the radar images, which are also shown in these figures. Lighting strikes are always recorded in the surroundings or the position of the considered anemometer and around the time of maximum wind speed. While during a thunderstorm, lightning strikes and high reflectivity values always occur within the cumulonimbus cloud, the gust front position associated with the thunderstorm itself can also be outside and slightly shifted in time with respect to the area of maximum convection and lightning activity. The first event (Fig. 11) occurred in summer, late in the afternoon, when diffuse unstable atmospheric conditions in the Padan Plane were favorable to the development of local deep convection in the western part of the Alps. Similar diffuse unstable conditions occurred in the afternoon in Fig. 12, which shows some scattered convective activity all over the Apennines in central Italy as well as in the northeastern Italy and Padan Plane, and more consistent cloud cover in western Italy. Figs. 13-15, conversely, correspond to the passage of frontal zones coming from south or southwestern, strongly interacting with the complex orography of the Alps and Apennines. 

In particular, the figures show the mean velocity over a 1-hour period (horizontal dashed line), the peak velocity averaged over 1 sec (circle), and the position of each port on the radar image. From the figures, it can be seen that the shapelet based classifier was able to detect a large variety of thunderstorms with different duration and wind speeds, allowing it to generalize well to thunderstorms that are not similar to the ones in the training set. Also, it can be seen from Fig.15 that even if the fixed window size of 1-hour does not capture the complete thunderstorm event (like in Fig. 11-14), the algorithm is able to accurately detect and label it as “Thunderstorm”. Once the detections are complete, a post-processing module extracts the value of peak velocity in the 1-hour segment and its exact time of occurrence. This timestamp is stored in the thunderstorm database, which can then be used to query and view the complete thunderstorm time series if needed. Fig. 16 shows the 1-s peak wind velocity values of the newly discovered non-stationary records divided into classes of membership for each port. Interestingly, many of the records have a peak velocity between 10-15 m/s.

\begin{figure}[ht]
  \centering
  \captionsetup{justification=centering}
  \includegraphics[scale=0.64]{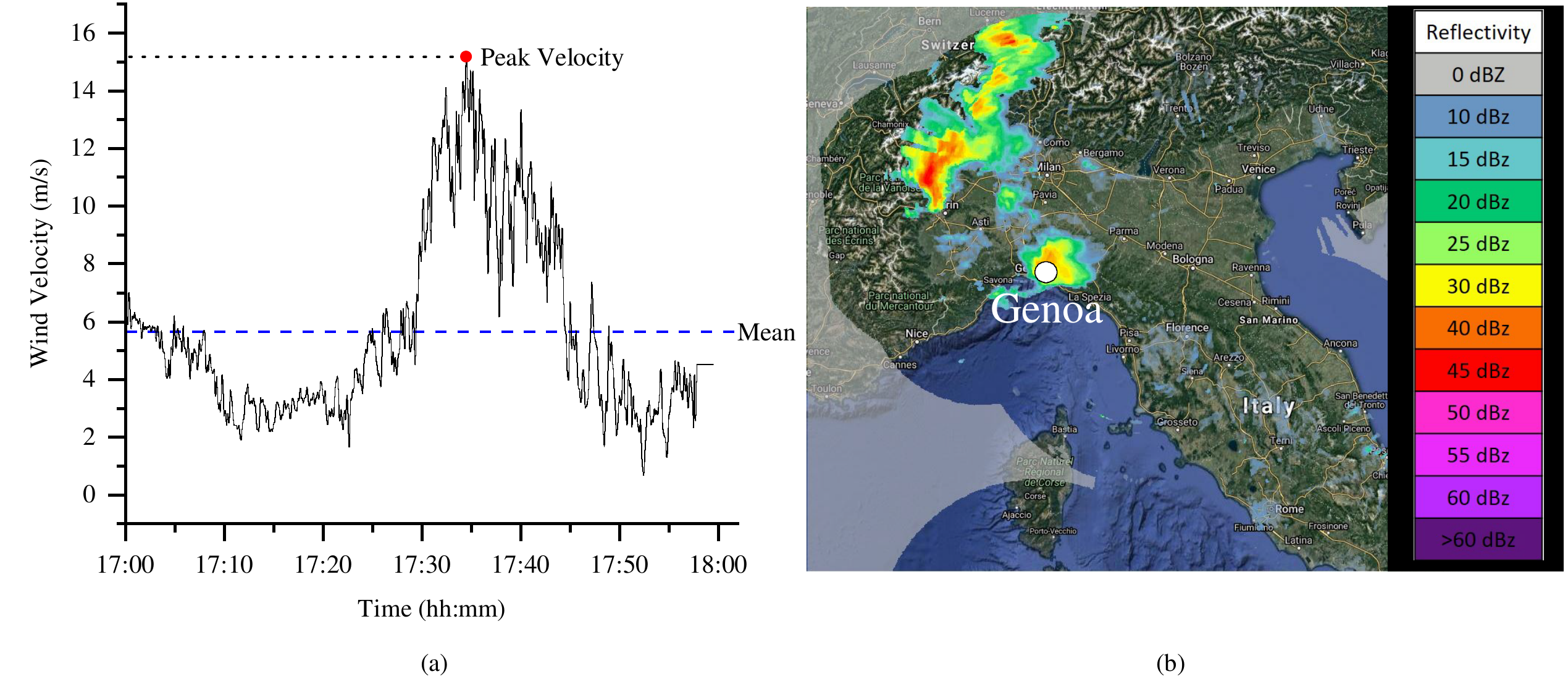}
  \caption{True Positive: Thunderstorm outflow recorded by Anemometer 1, Port of Genoa, 29 August 2012 (a) wind velocity in a 1-hour period (b) Vertical Maximum Intensity (VMI) from radar reflectivity displaying the thunderstorm.}
  \label{fig:fig11}
\end{figure}

\begin{figure}[htbp]
  \centering
  \captionsetup{justification=centering}
  \includegraphics[scale=0.64]{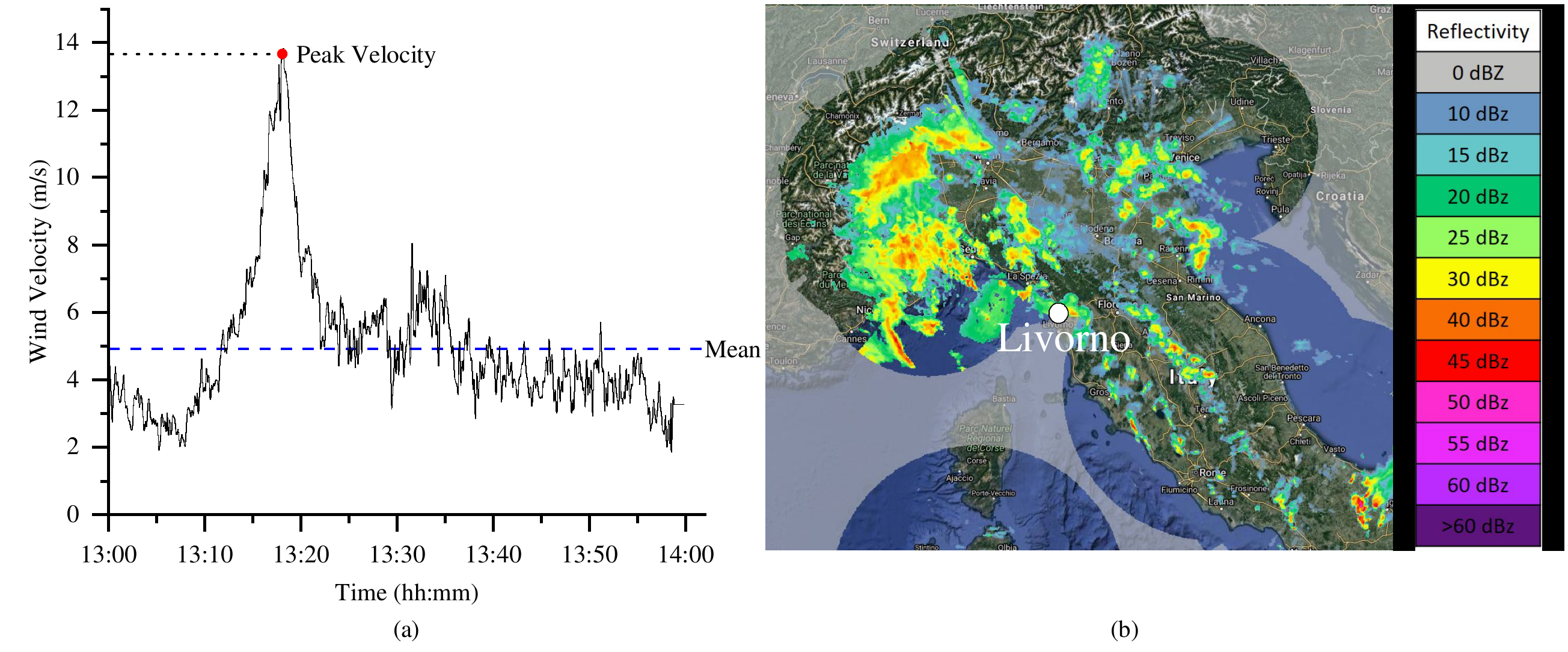}
  \caption{True Positive: Thunderstorm outflow recorded by Anemometer 4, Port of Livorno, 2 October 2015 (a) wind velocity in a 1-hour period (b) Vertical Maximum Intensity (VMI) from radar reflectivity displaying the thunderstorm.}
  \label{fig:fig12}
\end{figure}

\begin{figure}[htbp]
  \centering
  \captionsetup{justification=centering}
  \includegraphics[scale=0.64]{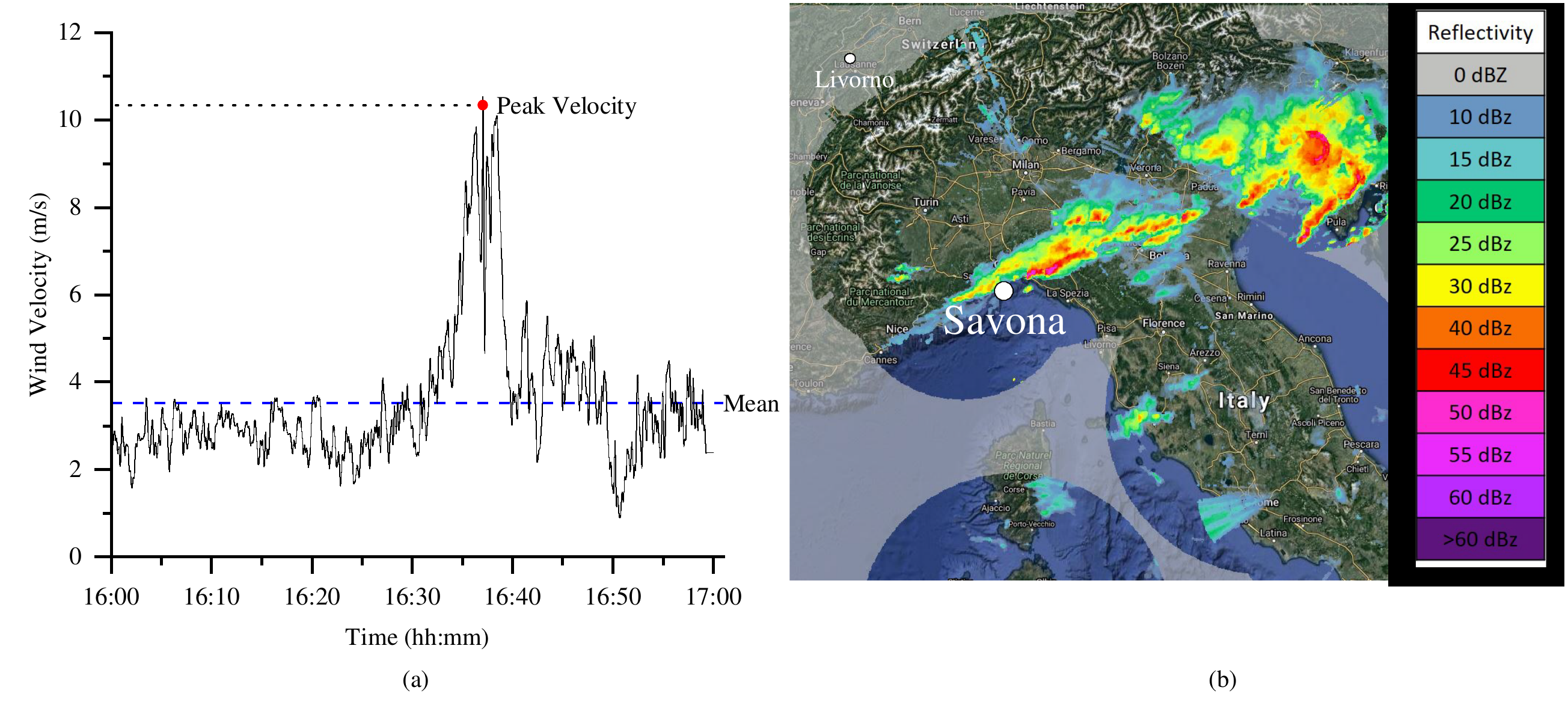}
  \caption{True Positive: Thunderstorm outflow recorded by Anemometer 1, Port of Savona, 23 June 2015 (a) wind velocity in a 1-hour period (b) Vertical Maximum Intensity (VMI) from radar reflectivity displaying the thunderstorm.}
  \label{fig:fig13}
\end{figure}

\begin{figure}[htbp]
  \centering
  \captionsetup{justification=centering}
  \includegraphics[scale=0.64]{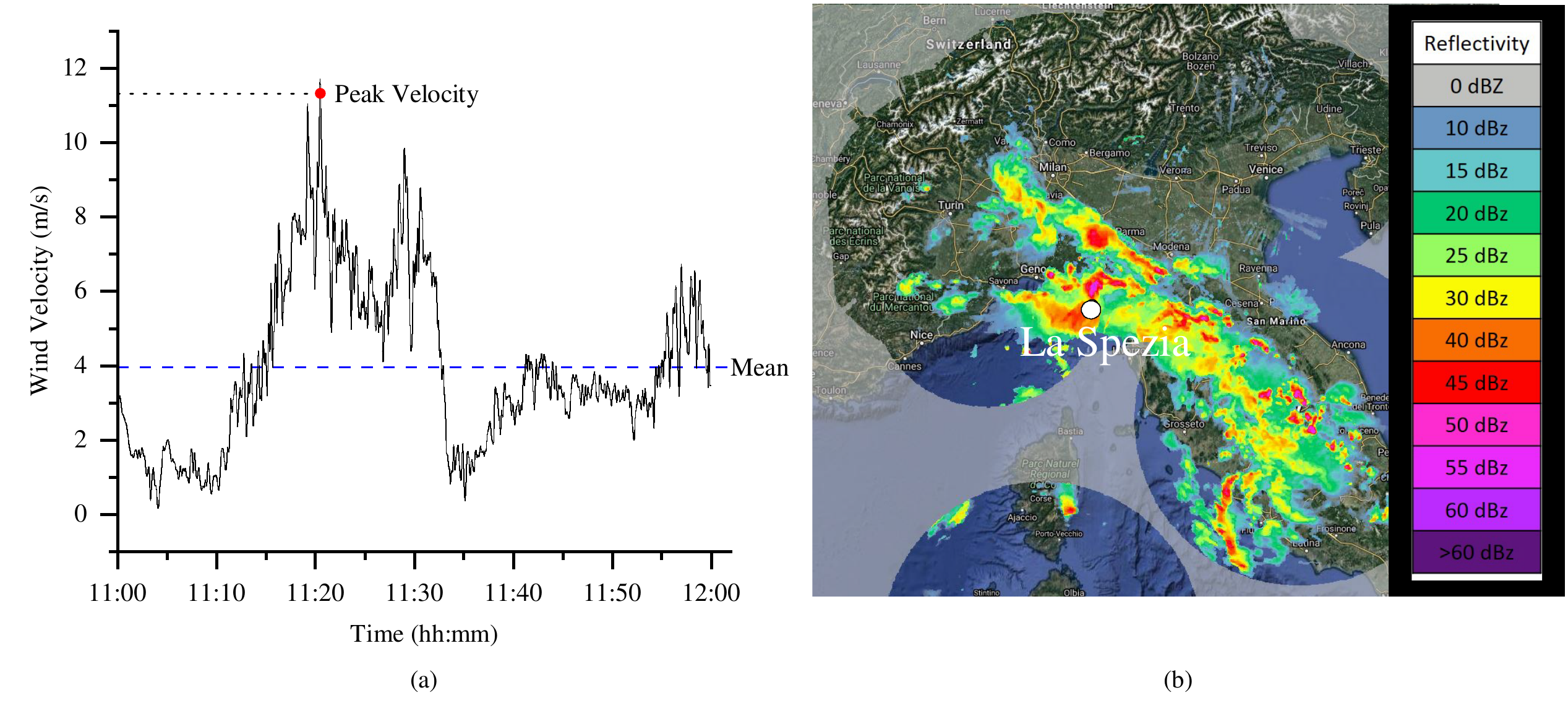}
  \caption{True Positive: Thunderstorm outflow recorded by Anemometer 3, Port of La Spezia, 10 August 2015 (a) wind velocity in a 1-hour period (b) Vertical Maximum Intensity (VMI) from radar reflectivity displaying the thunderstorm.}
  \label{fig:fig14}
\end{figure}

\begin{figure}[htbp]
  \centering
  \captionsetup{justification=centering}
  \includegraphics[scale=0.7]{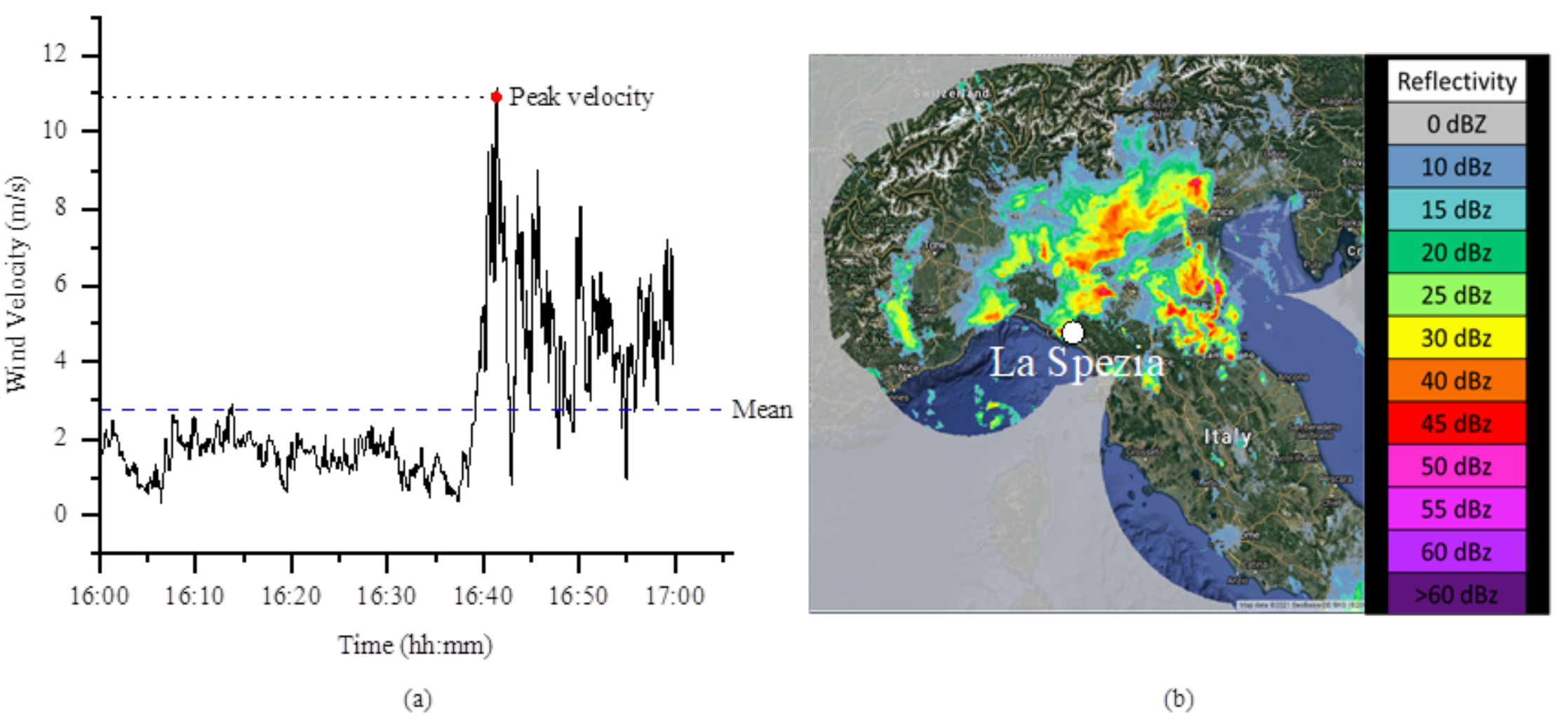}
  \caption{True Positive: Thunderstorm outflow recorded by Anemometer 3, Port of La Spezia, 21 May 2015 (a) wind velocity in a 1-hour period (b) Vertical Maximum Intensity (VMI) from radar reflectivity displaying the thunderstorm}
  \label{fig:fig15}
\end{figure}

\begin{figure}[htbp]
  \centering
  \captionsetup{justification=centering}
  \includegraphics[scale=0.74]{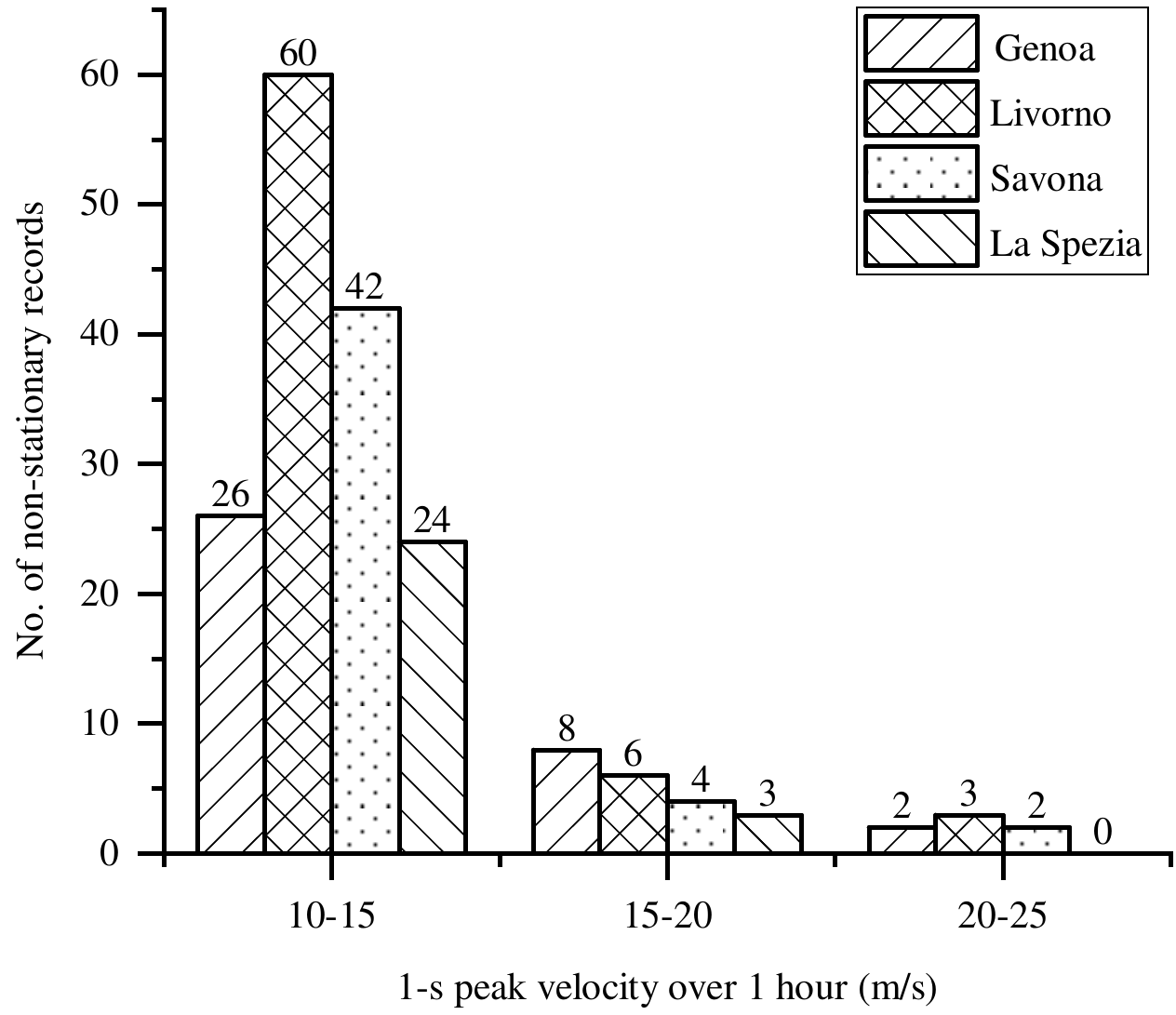}
  \caption{Distribution of peak wind velocities of newly discovered non-stationary records for each port}
  \label{fig:fig16}
\end{figure}

\begin{figure}[htbp]
  \centering
  \captionsetup{justification=centering}
  \includegraphics[scale=0.64]{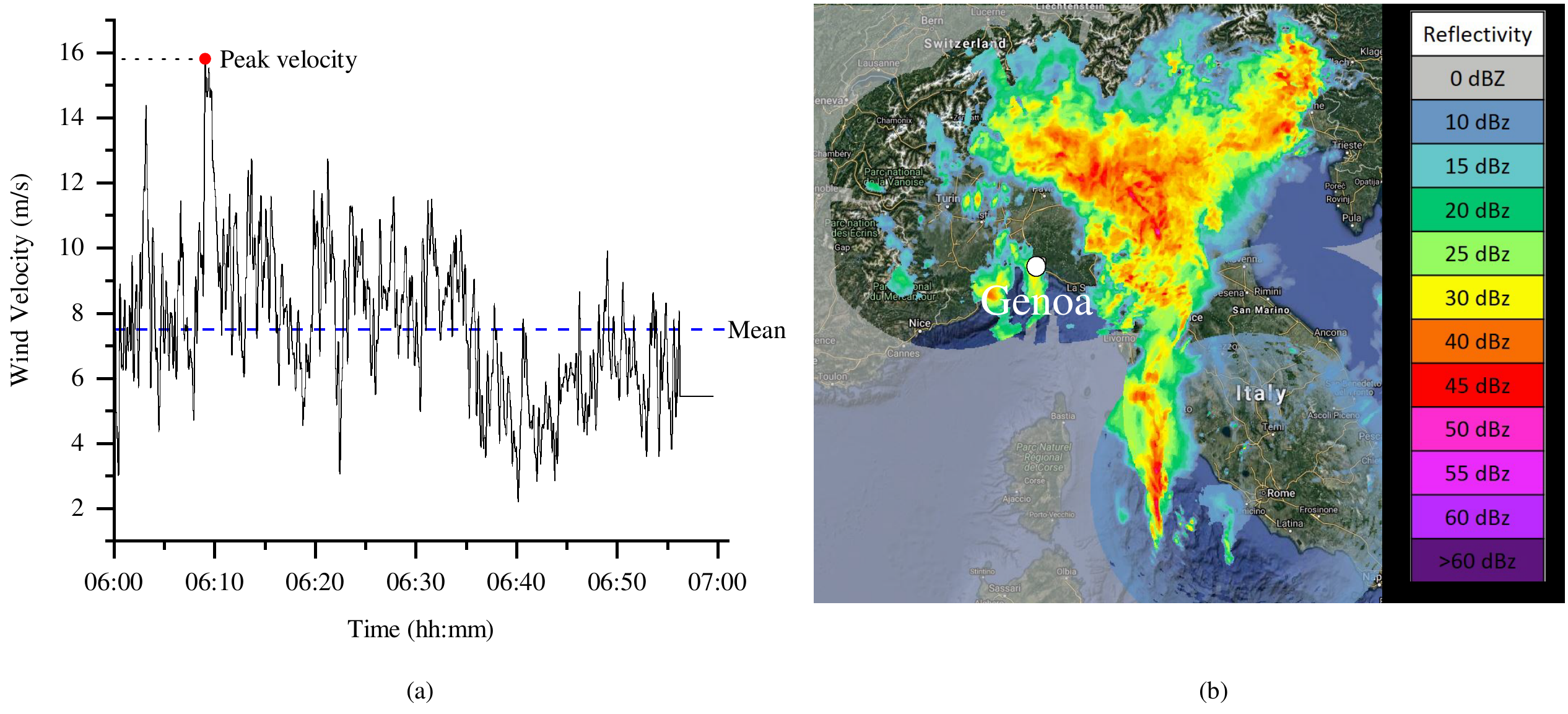}
  \caption{False-negative: Thunderstorm outflow recorded by Anemometer 1, Port of Genoa, 11 November 2012 (a) wind velocity in a 1-hour period (b) Vertical Maximum Intensity (VMI) from radar reflectivity displaying the thunderstorm}
  \label{fig:fig17}
\end{figure}

\begin{figure}[htbp]
  \centering
  \captionsetup{justification=centering}
  \includegraphics[scale=0.64]{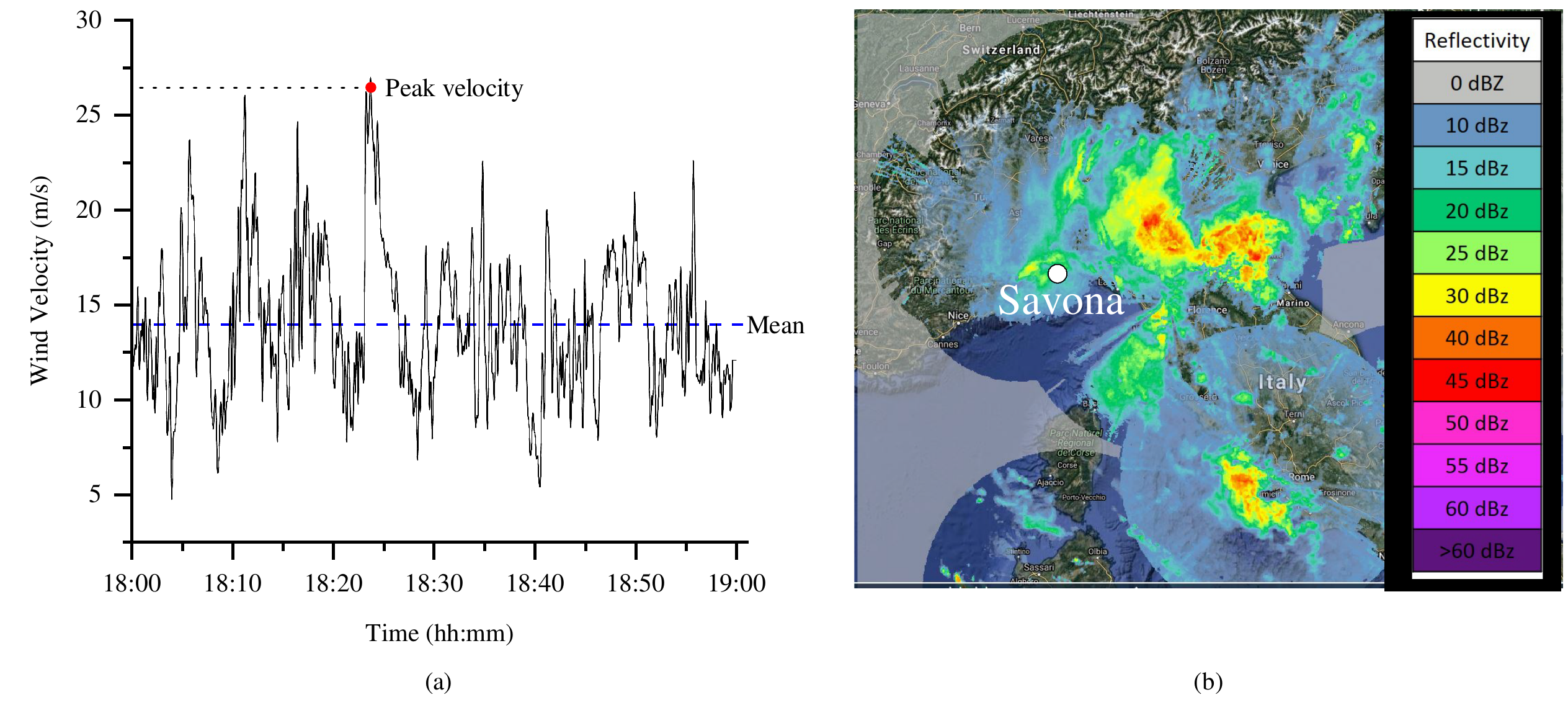}
  \caption{False-negative: Thunderstorm outflow recorded by Anemometer 4, Port of Savona, 24 February 2015 (a) wind velocity in a 1-hour period (b) Vertical Maximum Intensity (VMI) from radar reflectivity displaying the thunderstorm}
  \label{fig:fig18}
\end{figure}

\begin{figure}[htbp]
  \centering
  \captionsetup{justification=centering}
  \includegraphics[scale=0.64]{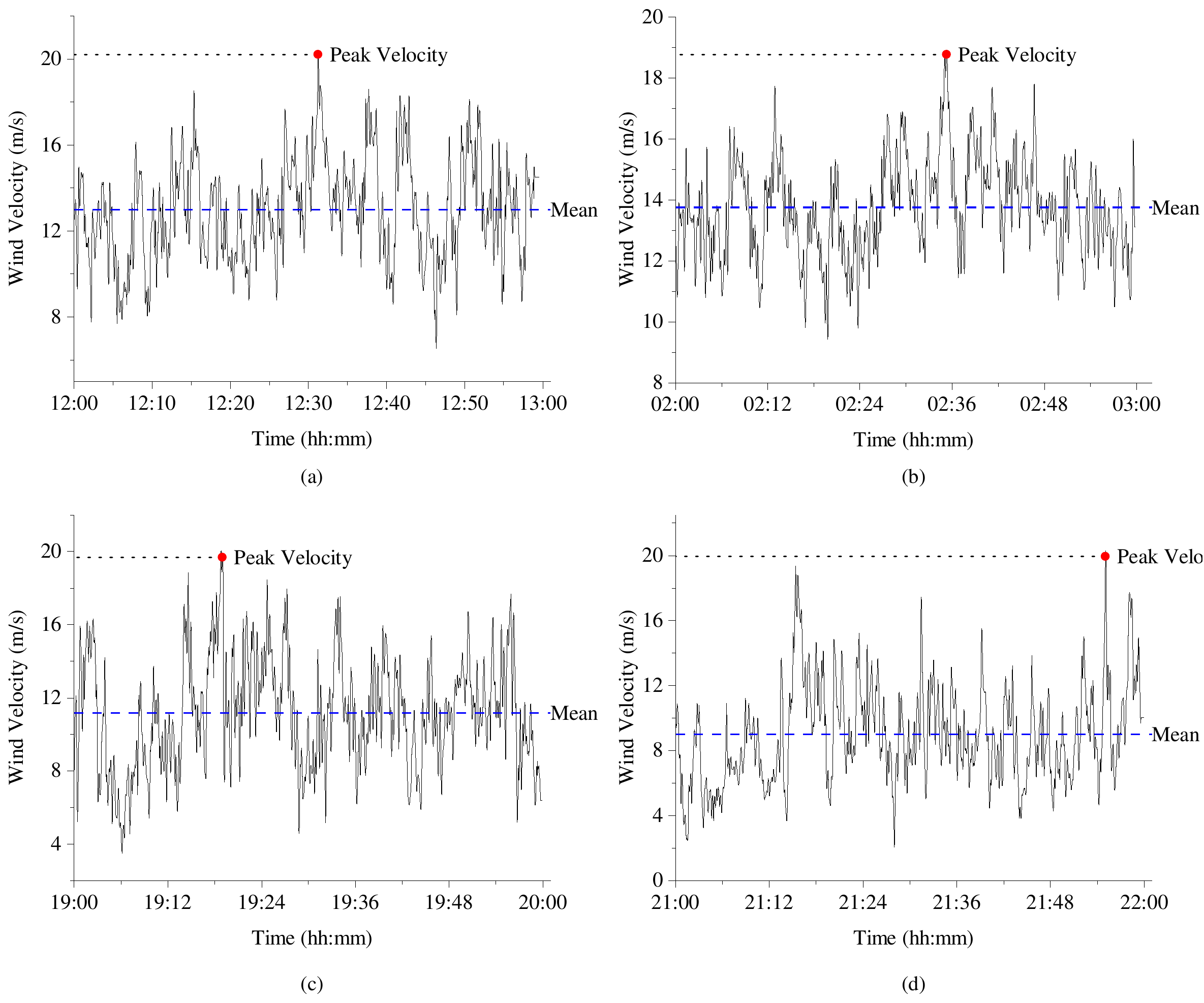}
  \caption{False-positives: Wind velocity in a 1-hour period recorded by (a) Anemometer 1, Port of Genoa, 8 February 2012 (b) Anemometer 4, Port of Livorno, 15 October 2015 (c) Anemometer 5, Port of Savona, 24 February 2015 (d) Anemometer 3, Port of La Spezia, 29 January 2015}
  \label{fig:fig19}
\end{figure}

From Table 3, it can also be seen that a total of 34 false negatives were obtained, i.e., thunderstorms that were missed by the shapelet algorithm. A few examples of these false negatives are shown in Fig.17 and 18. Compared to Figs. 11-15, the 1-hr non-stationary records in Fig.17 and 18 have peaks with a very short ramp up and ramp down duration, thus making it a challenging task to identify these small shapes. However, these shapes appear more significantly in a 10-min time window \cite{burlando2018monitoring}. One way to overcome this difficulty is allowing the shapelet algorithm to learn on 10-min wind velocity datasets instead of 1-hr. But this will lead to a deluge of datasets, and the thunderstorms with a longer duration cannot be effectively captured using this time window.

Also, a total of 1605 events were that were actually “other” were classified as “thunderstorms”. Fig. 19 shows examples of false positive records from each port. A significant portion of the false positive records was stationary, non-Gaussian events with relatively small mean wind velocities, large and repeated peaks, and moderately high gust factors, as seen from the figures. These are herein referred to as intermediate events \cite{kasperski2002new,zhang2018refined}. Until a systematic meteorological survey and interpretation of these events are carried out, it is reasonable to advance the hypothesis that they are associated with strongly unstable atmospheric conditions, or downslope winds, or recirculating vortices as well \cite{burlando2018monitoring}. The presence of a third class of wind events that look similar to thunderstorms in terms of large peaks but with intermediate properties greatly challenges the binary classification problem. It can be seen from Table 3 that the false positives significantly affect the classification accuracy of data from certain anemometers. A possible strategy to overcome this challenge is to transform the 2-class problem (Thunder vs Other) into a 3-class problem (Thunder vs Other vs Intermediate), as explained in the following section.

\section{MULTI-CLASS CLASSIFICATION}
As mentioned in Sections 5.1.1 and 5.1.2, the learning set for binary classification contains 120 thunderstorm records labeled as “Thunderstorms” and 120 non-thunderstorm records labeled as “Other”. The thunderstorms records are obtained from the Port of Livorno as it is well known to experience frequent thunderstorms. For multi-class classification, 120 intermediate events recorded at the Port of Livorno are labeled as “Intermediate” and added to the existing learning set. So the learning set now contains three classes: Thunderstorms, Other and Intermediate. The methodology explained in Section 5 is then repeated. The learning set is randomly split into training (70\%) and test (30\%) sets. Thus the training set has 252 time series, and the testing set has 108 time series. The time series in the training set is used to discover shapelets. For the multi-class classification problem, a total of 154 shapelets is discovered from the training set using the shapelet algorithm. The first six shapelets (highlighted in red) with the highest IG are shown in Fig.20. Shapelets 1-3 are from intermediate events, and shapelets 4-6 are from thunderstorm time series. Out of the 154 shapelets, 34 correspond to non-thunderstorm records, and 52 correspond to intermediate events. To train the machine learning algorithm for thunderstorm identification, the time series in the training set is transformed to a local shape space using the 154 discovered shapelets. Thus a 252 x 154 matrix is constructed where each element is the minimum Euclidean distance between a time series and a shapelet in the training set. Similarly, the testing set is also transformed using the discovered shapelets, and a 108 x 32 matrix is constructed. As shown in stage 2 of the methodology in Fig.4, the Random Forest classifier (500 trees) is first trained on the testing set. The trained algorithm is then used to detect thunderstorms from all anemometers in the year 2015, and the results are discussed in the next section. 

\begin{figure}[htbp]
  \centering
  \captionsetup{justification=centering}
  \includegraphics[scale=0.64]{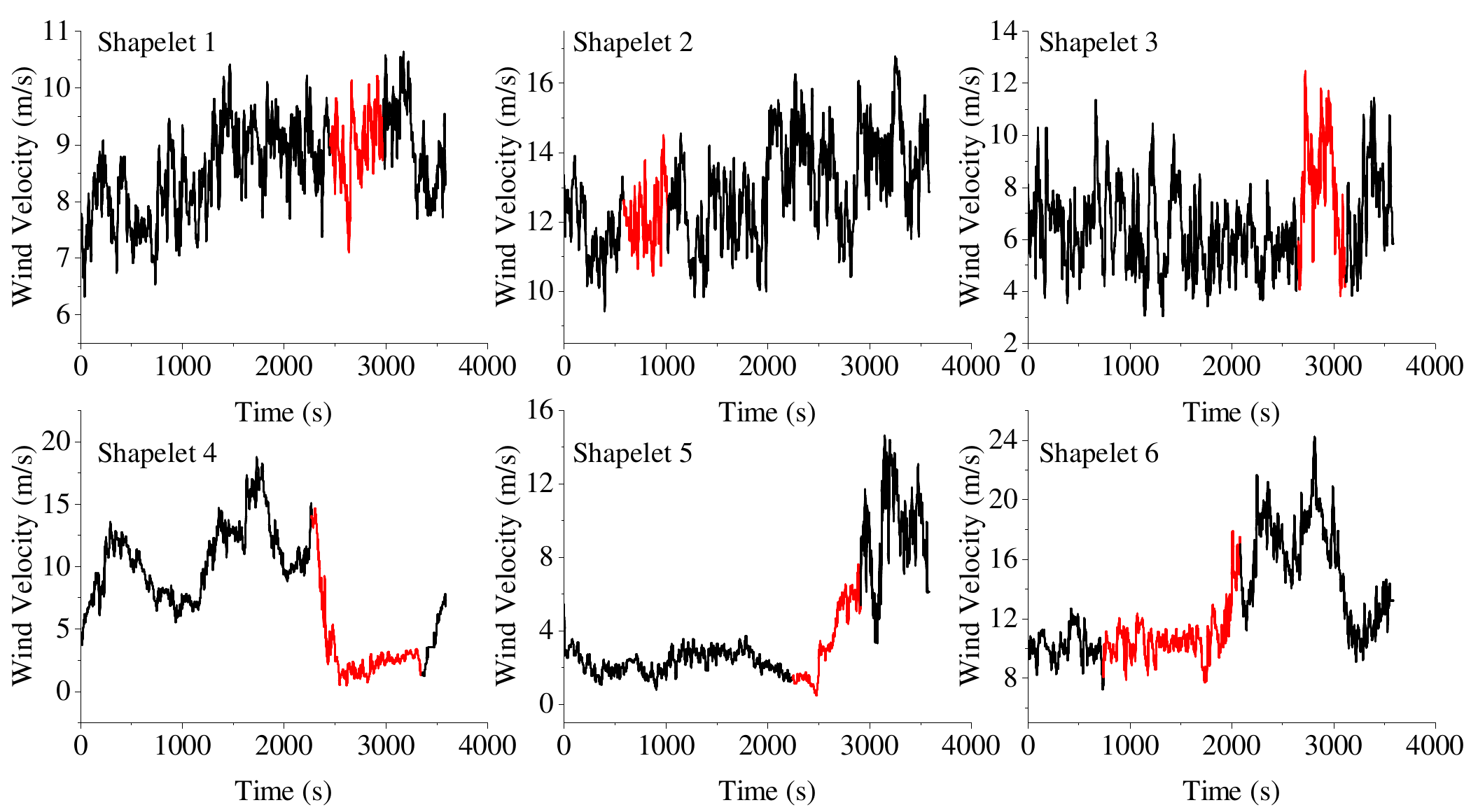}
  \caption{First six shapelets (highlighted in red) discovered for the 3-class classification problem}
  \label{fig:fig20}
\end{figure}

\begin{figure}[htbp]
  \centering
  \captionsetup{justification=centering}
  \includegraphics[scale=0.23]{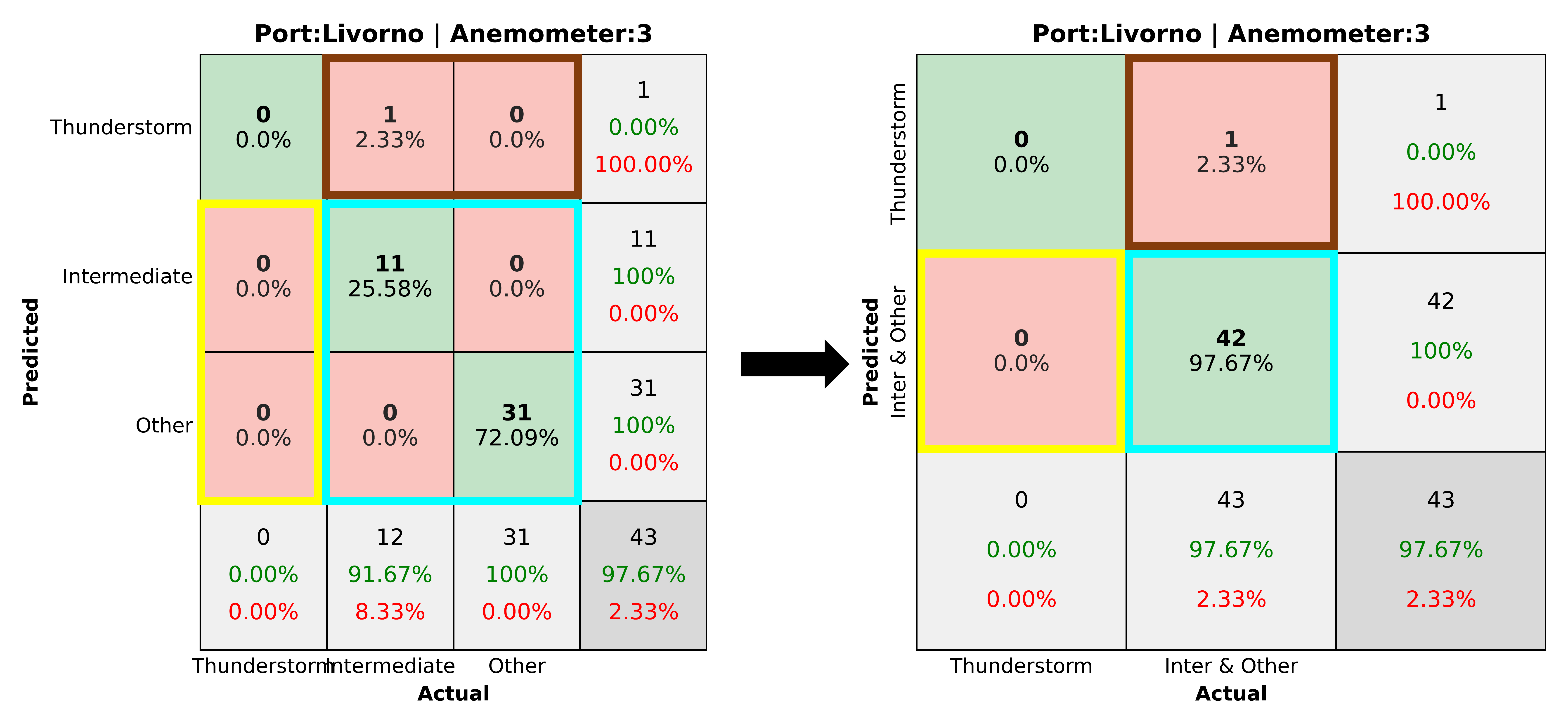}
  \caption{Concise confusion matrix}
  \label{fig:fig21}
\end{figure}

\begin{figure}[htbp]
  \centering
  \captionsetup{justification=centering}
  \includegraphics[scale=0.36]{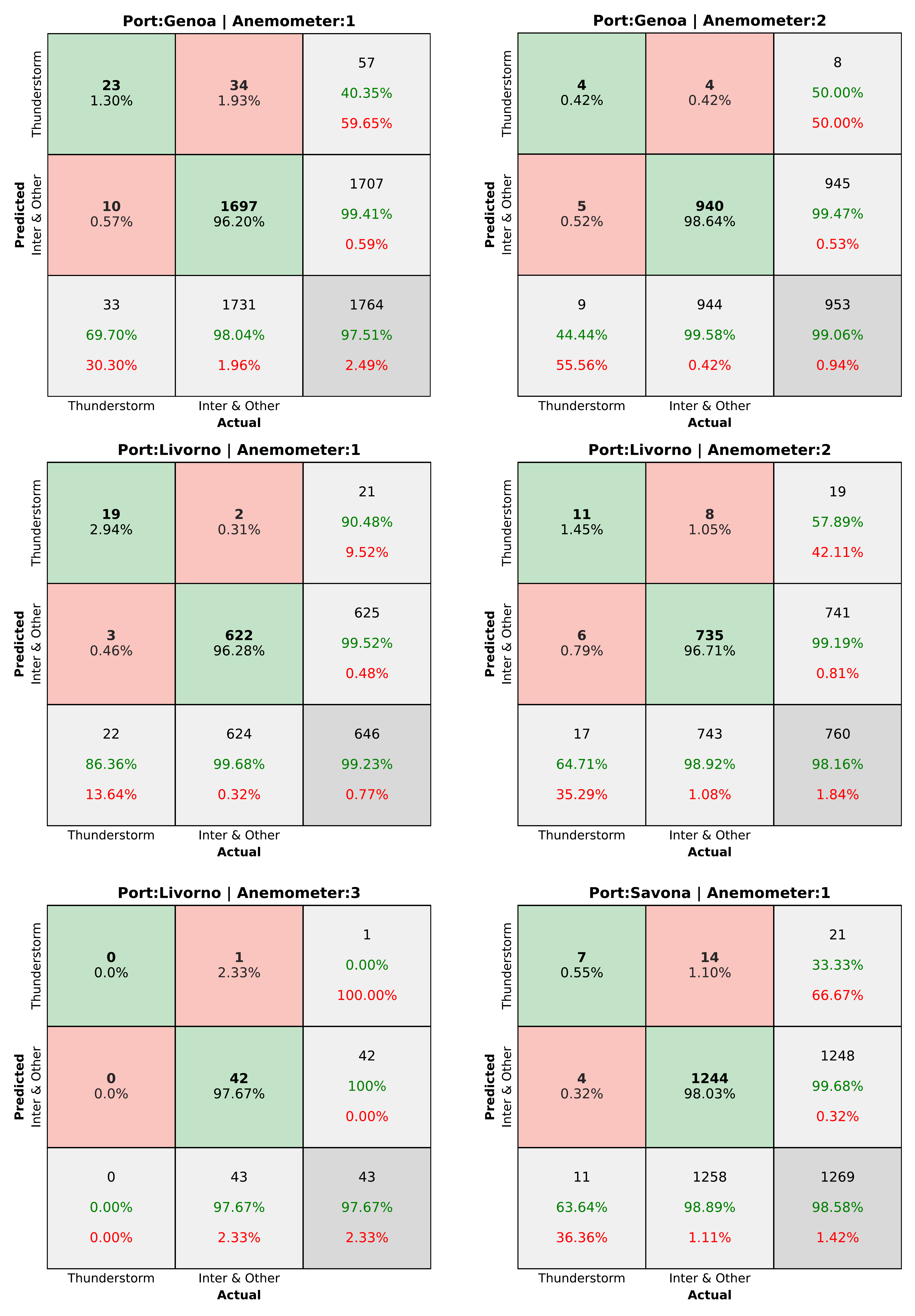}
  \caption{Concise confusion matrices for Port of Genoa, Livorno and Savona (Anemometer:1)}
  \label{fig:fig22}
\end{figure}

\begin{figure}[htbp]
  \centering
  \captionsetup{justification=centering}
  \includegraphics[scale=0.36]{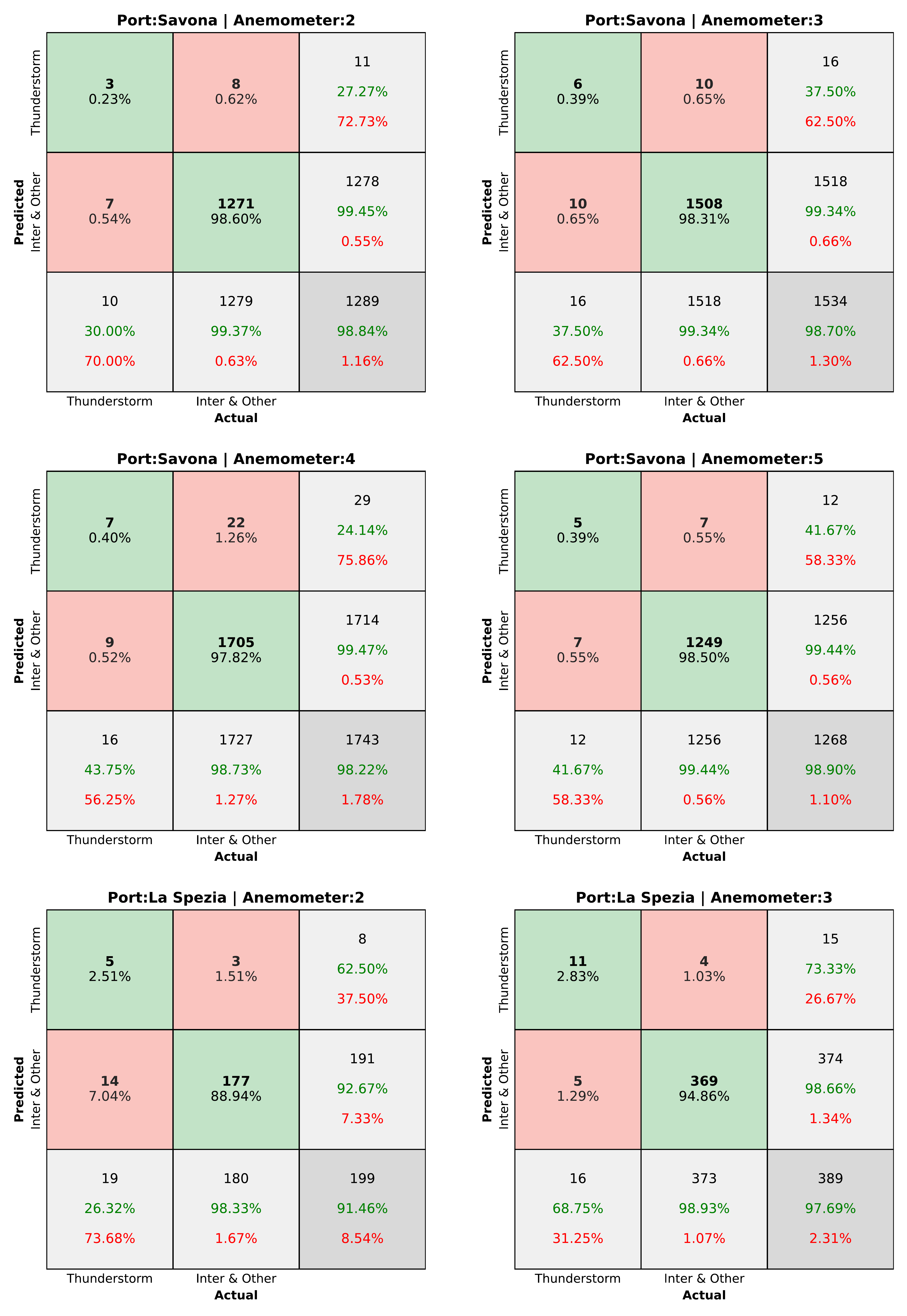}
  \caption{Concise confusion matrices for Port of Savona and La Spezia}
  \label{fig:fig23}
\end{figure}

\begin{table}
\centering
\captionof{table}{Comparison of key performance metrics between binary and multi-class classification }
\begin{tabular}{ccccccccc} 
\hline
\multirow{2}{*}{Port}                                                          & \multirow{2}{*}{Year} & \multirow{2}{*}{Anemometer} & \multicolumn{2}{c}{False Positive (FP)} & \multirow{2}{*}{\begin{tabular}[c]{@{}c@{}}\% Decrease \\ in FP\end{tabular}} & \multicolumn{2}{c}{Overall Accuracy} & \multirow{2}{*}{\begin{tabular}[c]{@{}c@{}}\% Increase \\ in accuracy\end{tabular}}  \\ 
\cline{4-5}\cline{7-8}
                                                                               &                       &                             & Binary & Multi-class                    &                                                                               & Binary & Multi-class                 &                                                                                      \\ 
\hline
\multirow{2}{*}{Genoa}                                                         & 2012                  & 1                           & 362    & 34                             & 91                                                                            & 0.80   & 0.98                        & 22.5                                                                                 \\
                                                                               & 2015                  & 2                           & 49     & 4                              & 92                                                                            & 0.96   & 0.99                        & 3.1                                                                                  \\
\multirow{4}{*}{Livorno}                                                       & \multirow{4}{*}{2015} & 1                           & 51     & 2                              & 96                                                                            & 0.93   & 0.99                        & 6.5                                                                                  \\
                                                                               &                       & 2                           & 53     & 8                              & 85                                                                            & 0.94   & 0.98                        & 4.3                                                                                  \\
                                                                               &                       & 3                           & 7      & 1                              & 86                                                                            & 0.85   & 0.98                        & 15.3                                                                                 \\
                                                                               &                       & 4                           & 109    & *N/A                           & *N/A                                                                          & 0.83   & *N/A                        & *N/A                                                                                 \\
\multirow{5}{*}{\begin{tabular}[c]{@{}c@{}}Savona \\ and \\ Vado\end{tabular}} & \multirow{5}{*}{2015} & 1                           & 141    & 14                             & 90                                                                            & 0.90   & 0.99                        & 10                                                                                   \\
                                                                               &                       & 2                           & 143    & 8                              & 94                                                                            & 0.90   & 0.99                        & 10                                                                                   \\
                                                                               &                       & 3                           & 238    & 10                             & 96                                                                            & 0.85   & 0.99                        & 16.5                                                                                 \\
                                                                               &                       & 4                           & 228    & 22                             & 90                                                                            & 0.88   & 0.98                        & 11.4                                                                                 \\
                                                                               &                       & 5                           & 113    & 7                              & 94                                                                            & 0.92   & 0.99                        & 7.6                                                                                  \\
\multirow{2}{*}{La Spezia}                                                     & \multirow{2}{*}{2015} & 2                           & 25     & 3                              & 88                                                                            & 0.86   & 0.92                        & 7                                                                                    \\
                                                                               &                       & 3                           & 86     & 4                              & 95                                                                            & 0.80   & 0.98                        & 22.5                                                                                 \\ 
\hline
\multicolumn{9}{c}{\begin{tabular}[c]{@{}c@{}}* Intermediate events obtained from this anemometer are used for training the algorithm \\ and hence classification is not performed on these datasets\end{tabular}}                                                                                                                                                                          
\end{tabular}
\end{table}

\subsection{Results and Discussion}
The Shapelet-based Random Forest classifier is used to detect thunderstorms records from all the ultrasonic anemometers except anemometer no.4 at the Port of Livorno. This is because, as mentioned in the previous section,  intermediate events are obtained from this anemometer for training the algorithm, and hence classification is not performed on these datasets. Since there are three classes, a summary of the performance metrics of the classifier is shown visually in the form of a confusion matrix rather than as a table, as shown in Fig. 21. The rows correspond to the predicted class, and the columns correspond to the actual class. The diagonal cells correspond to observations that are correctly classified. The off-diagonal cells correspond to incorrectly classified observations. Both the number of observations and the percentage of the total number of observations are shown in each cell. The column on the far right of the matrix shows the percentages of all the examples predicted to belong to each class that are correctly and incorrectly classified. These metrics are often called the precision (or positive predictive value) and false discovery rate, respectively. The row at the bottom of the plot shows the percentages of all the examples belonging to each class that are correctly and incorrectly classified. These metrics are often called the recall (or true positive rate) and false negative rate, respectively. The cell in the bottom right of the plot shows the overall accuracy. Since the present study is focused on identifying thunderstorms, the original confusion matrix is transformed into a concise confusion matrix, as shown in Fig.21, to easily interpret the results.  True Positive (TP) is the same for both original and concise confusion matrix. True Negative (TN) in the concise matrix is the sum of all the elements highlighted in cyan in the original confusion matrix. False Positive (FP) is the sum of all the elements highlighted in brown in the original confusion matrix and False Negative (FN) is the sum of all the elements highlighted in yellow in the original confusion matrix.

The concise confusion matrix for all the ports is shown in Fig. 22 and Fig. 23. Consider anemometer no.1 at the Port of Livorno. In this confusion matrix, the first two diagonal cells show the number and percentage of correct classifications. For example, 19 time series are correctly classified as thunderstorms. This corresponds to 2.94\% of all time series records. Similarly, 622 time series are correctly classified as either other or intermediate. This corresponds to 96.28\% of all 646 time series records. 2 time series are incorrectly classified as thunderstorm and this corresponds to 0.31\% of all time series records. Similarly, 3 time series are incorrectly classified as either other or intermediate, and this corresponds to 0.46\% of all data. Out of 21 thunderstorm predictions, 90.48\% are correct, and 9.52\% are wrong. Out of 625 predictions as either other or intermediate, 99.52\% are correct, and 0.48\% are wrong. Out of 22 time series corresponding to thunderstorms, 86.36\% are correctly predicted as thunderstorms, and 13.64\% are predicted as either other or intermediate. Out of 624 time series that are either other or intermediate, 99.68\% are correctly classified, and 0.32\% are classified as thunderstorms. Overall, 99.23\% of the predictions are correct, and 0.77\% are wrong. The rest of the confusion matrices can be interpreted the same way. A comparison of key performance metrics between binary and multi-class classification for all the anemometers is provided in Table 4. It can be seen that the number of false positives has reduced significantly for the multi-class classification when compared to the binary classification. An average of 92\% reduction is seen in the number of false positives. Moreover, the overall accuracy of the classification has also increased by an average of 11\% for the multi-class classification.

\section{CONCLUDING REMARKS}
This paper proposes a new avenue of research that uses machine learning techniques, independent of wind characteristics-based parameters, to autonomously identify and separate desired wind events, thunderstorms in the present case, from large volumes of continuous data. This would complement the meteorological and wind engineering methodologies to identify thunderstorms. This objective is achieved using a relatively new and efficient time series representation named “Shapelet transform”  couched in a machine learning algorithm (Random Forest classifier) to identify thunderstorms from anemometric records. The Shapelet Transform is a unique time series representation technique that is solely based on the shape of the time series and provides a universal standard feature for detection, which is based on the Euclidean distance between a shapelet and a time series. This method is designed to be used to detect thunderstorms in large data sets with continuously recorded data.  In this study, the methodology is used to identify thunderstorms from 1-year of data from 14 ultrasonic anemometers that are a part of the “Wind and Ports” and “Wind, Ports and Sea” project – an extensive in-situ monitoring network aimed at investigating extreme wind events in port areas of Italy. A total of 240 non-stationary records associated with a wide variety of thunderstorms with varying duration and wind velocities were identified using the shapelet transform method. The results lead to a comprehensive understanding of a wide variety of thunderstorms that have not been previously detected using conventional gust factor-based methods.

\section*{ACKNOWLEDGEMENTS}

This work was supported in part by the Robert M. Moran Professorship and National Science Foundation Grant (CMMI 1612843). The contribution of the third and fourth authors is funded by the European Research Council (ERC) under the European Union’s Horizon 2020 research and innovation program (grant agreement No. 741273) for the project "THUNDERR - Detection, simulation, modeling and loading of thunderstorm outflows to design wind-safer and cost-efficient structures" – through an Advanced Grant 2016. The data used for this research were recorded by the monitoring network set up as part of the European Projects “Winds and Ports” (grant No. B87E09000000007) and “Wind, Ports and Sea” (grant No. B82F13000100005), funded by the European Territorial Cooperation Objective, Cross-border program Italy-France Maritime 2007–2013. This paper is dedicated to the fond memory of Prof. Giovanni Solari, who sadly passed away during the manuscript preparation and the review process. He will be remembered for his dedication and contributions to thunderstorms and their impact on structures.

\bibliographystyle{unsrt}  
\bibliography{references}  %%% Remove comment to use the external .bib file (using bibtex).


% \bibitem{allen1978automatic}
% George Kour and Raid Saabne.
% \newblock Fast classification of handwritten on-line arabic characters.
% \newblock In {\em Soft Computing and Pattern Recognition (SoCPaR), 2014 6th
%   International Conference of}, pages 312--318. IEEE, 2014.

% \bibitem{hadash2018estimate}
% Guy Hadash, Einat Kermany, Boaz Carmeli, Ofer Lavi, George Kour, and Alon
% \newblock Estimate and replace: A novel approach to integrating deep neural
%   networks with existing applications.
% \newblock {\em arXiv preprint arXiv:1804.09028}, 2018.

% \end{thebibliography}

\end{document}